\documentclass[referee]{raa}
\usepackage{graphicx,times}
\usepackage{natbib}
\usepackage{amssymb,amsmath}
\usepackage{cite}
\bibpunct{(}{)}{;}{a}{}{,}

\usepackage[a4paper=true,dvipdfm=true,pagebackref=true]{hyperref}
\hypersetup{pdftitle = The title of my PDF, pdfauthor = My name, pdfsubject= The subject, pdfkeywords = keyword1 keyword2 keyword3}
\hypersetup{colorlinks = true, linkcolor = green, anchorcolor = red, citecolor = blue, filecolor = red, pagecolor = red, urlcolor = red}

\begin{document}

   \title{The Point Spread Function Reconstruction by Using Moffatlets -- \uppercase\expandafter{\romannumeral1} $^*$
\footnotetext{\small $*$ Supported by the National Natural Science Foundation of China.}
}

 \volnopage{ {\bf 2012} Vol.\ {\bf X} No. {\bf XX}, 000--000}
   \setcounter{page}{1}

   \author{Baishun Li\inst{1,2}, Guoliang Li\inst{1}, Jun Cheng\inst{3}, John Peterson\inst{3}, Wei Cui\inst{3}
   }

   \institute{ Purple Mountain Observatory, Chinese Academy of Sciences, Nanjing,210000,
China; {\it libaishun@pmo.ac.cn}\\
        \and
             University of Chinese Academy of Sciences,
             Beijing, China\\
             \and
             Department of Physics, Purdue University, 525 Northwestern Ave, West Lafayette, Indiana 47907, UAS\\
\vs \no
  % {\small Received 2012 June 12; accepted 2012 July 27}
}

\abstract{
The shear measurement is a crucial task in the current and the future weak lensing survey projects. And the reconstruction of the point spread function(PSF) is one of the essential steps. In this work, we present three different methods, including Gaussianlets, Moffatlets and EMPCA to quantify their efficiency on PSF reconstruction using four sets of simulated LSST star images. Gaussianlets and Moffatlets are two different sets of basis functions whose profiles are based on Gaussian and Moffat functions respectively. Expectation Maximization(EM) PCA is a statistical method performing iterative procedure to find principal components of an ensemble of star images. Our tests show that: 1) Moffatlets always perform better than Gaussianlets. 2) EMPCA is more compact and flexible, but the noise existing in the Principal Components (PCs) will contaminate the size and ellipticity of PSF while Moffatlets keeps them very well.
}

   \authorrunning{Baishun Li,Guoliang Li,Jun Cheng,Wei Cui }            %author_head in even pages
   \titlerunning{Modeling PSFs by using Gaussianlets, Moffatlets and EMPCA}  % title_head in odd pages
   \maketitle

%________________________________________________ sections below
%
\section{Introduction}           %% first-level sections will be auto-capitalized
\label{sect:intro}

Gravitational lensing provides a unique way to map the matter distribution in the Universe. By measuring the shape distortion of the distant galaxies, one can gain the lensing signals and thus study the mass distribution in clusters of galaxies (\citealt{1999ARA&A..37..127M}), large scale structures (\citealt{2003ARA&A..41..645R,2013MNRAS.433.3373V,2014A&A...571A..17P}), and probes directly the invisible dark sector and the fundamental nature of gravity (\citealt{2008ARNPS..58...99H,2010RPPh...73h6901M,2010GReGr..42.2177H,2006astro.ph..8675M}). However the shape of a galaxy can be distorted by several different mechanisms , such as  1)sheared by lensing effect. 2)convolved with a Point Spread Function (PSF). 3)pixelated on CCD and finally affected by noise. In order to recover accurately the original galaxy shape (shape right after the galaxy being lensed), decrease the statistic error and quantify the intrinsic  alignments of background galaxies (\citealt{2001PhR...340..291B}), a number of current and planned large-area surveys were proposed, such as Euclid (\citealt{2011arXiv1110.3193L}), LSST\footnote{{\color{blue}{\url{http://www.lsst.org}}}} (\citealt{2009arXiv0912.0201L}), WFIRST-AFTA(\citealt{2015arXiv150303757S}), to reduce the statistical uncertainty. On the other hand, a variety of weak lensing shear measurement algorithms (\citealt{1995ApJ...449..460K,1997ApJ...475...20L,1998ApJ...504..636H,2003MNRAS.338...48R}) have been proposed and a series of data analysis challenges, such as GREAT08 (\citealt{2010MNRAS.405.2044B}),GREAT10 (\citealt{2012MNRAS.423.3163K,2013ApJS..205...12K}) and the most recent one GREAT03\footnote{{\color{blue}{\url{http://great3challenge.info}}}}\footnote{{\color{blue}{\url{http://great3.projects.phys.ucl.ac.uk/leaderboard/}}}} (\citealt{2014ApJS..212....5M}), have been carried out to improve the precision and reduce systematic biases.

One of the crucial parts in reducing the systematic biases in shear measurement is modeling the point spread function (PSF) to adequate precision.The scatter and systematic bias on the size and ellipticity of the reconstructed PSF will introduce systematic bias to the shear measurement (\citealt{2008A&A...484...67P,2009A&A...500..647P,2013MNRAS.429..661M}). PSF is the spreading of light caused by various complex physical processes, such as diffraction by the aperture of the telescope, imperfect optics and tracking systems, temperature variations in the camera, vibrations, optical changes during telescope refocusing, and turbulence in the atmosphere (a concern for ground-based telescopes). This means that the PSF can not be represented by a simple explicit function form. Gaussian PSF was usually assumed to serve as a good approximation for most astronomical cases. But it deviates the real PSF due to the existence of ¡°wings¡± in stellar profiles. The Moffat function is shown to describe well the presence of wings (when the value of $\beta$ is taken properly) and contain the Gaussian function as a limiting case (when $\beta \rightarrow \infty$) (\citealt{2001MNRAS.328..977T}). However to reproduce PSF arriving the weak lensing precision, high-order correction is required. Based on a technique called Shapelets (\citealt{2003MNRAS.338...35R,2003MNRAS.338...48R,2005MNRAS.363..197M}), which decompose an object using a series of localised basis functions and compress the information of shape in a small number of expansion coefficients. In this paper, we use the Gaussian and Moffat function as primary profiles to create two basis function sets which we call Gaussianlets and Moffatlets (\citealt{2013IAUS..288..306L}) respectively to decompose the PSF.

The minimum number of 50 stars over which the PSF must be calibrated in order to control the systematic errors to a level similar to the statistical errors has been estimated for the future ambitious surveys (\citealt{2008A&A...484...67P}). With this ensemble of stars in an image, a set of Principal Components (PCs) can be solved via performing the statistical procedure called principal component analysis (PCA) . In \citealt{2012PASP..124.1015B}, a framework called Expectation Maximization (EM) PCA is introduced that extended the classical PCA to a form that can incorporate estimates of measurement variance while solving for the PCs. In this paper we use this method to find the PCs of PSF.

This article is organized as follows. In section \ref{sect:2} we describe the three methods, Gaussianlets, Moffatlets and EMPCA, and their algorithms for reconstructing PSFs. In section \ref{sect:3} we describe the simulate structure of the data we use. And in section \ref{sect:4} we perform the numerical tests of our three methods and compare their reconstruction efficiency. Finally we conclude by discussing the limitations and prospects of our algorithm in section \ref{sect:5}.

% Authors can give a citation as `\citealt{Michel+etal+1992}'.
% You may also use \cite, \citep and \citet for citation, and use Table~1
% or Figure~1 and so forth. Using \ref and \label for cross-references of
% Tables/Figures is a good way in adjusting/adding/removing text, tables or
% figures.
\section{Reconstruction Methods}
\label{sect:2}
\subsection{Gaussianlets and Moffatlets}
\label{sect:2.1}
The so-called Gaussianlets here,  is a reduced version of shapelets \citep{2005MNRAS.363..197M} where we only keep the basis functions with m = 0. The explicit mathematical formula of Gaussianlets is
\begin{equation}
   P_l(r)=\frac{1}{\sqrt{\pi\sigma_d^2}}e^{-\frac{r^2}{2\sigma_d^2}}L_l(\frac{r^2}{\sigma_d^2}),
\label{eq:1}
\end{equation}
The basis functions in Moffatlets are also circular symmetric and contain no angular components.
The  formula ( see Appendix \ref{app:1} for more details) is
\begin{equation}
   Q_l(r)=\sqrt{\frac{2\beta-1}{\pi r^2_d}}L_l[v(r)][1+(\frac{r}{r_d})^2]^{-\beta},
\label{eq:2}
\end{equation}
where
\begin{equation}
   v(r)=(\frac{1}{2\beta}-1)ln[1+(\frac{r}{r_d})^2]^{-2\beta}
\end{equation}

In both $(1)$ and $(2)$, $L_l(x)$  is the Laguerre polynomials:
\begin{equation}
   L_l(x)=\frac{e^x}{l!}\frac{d^l}{dx^l}(e^{-x}x^l)=\frac{1}{l!}(\frac{d}{dx}-1)^lx^l.
\label{eq:3}
\end{equation}
where $l$ runs from $0$ to $\infty$. Mathematically both sets of basis functions are orthogonalized and normalized in the following sense,
\begin{equation}
   \int_{0}^{+\infty}{P_l(r)P_m(r)dr}=\int_{0}^{+\infty}{Q_l(r)Q_m(r)dr}=\delta_{lm}.
\label{eq:4}
\end{equation}

There are free parameters ($\sigma_d$ in Gaussianlets and $(\beta , r_d)$ in Moffatlets) that are adjustable in the basis functions. Tuning their values we can change the size and steepness of the basis functions' radial variation. In the process of modeling, these free parameters are adjusted to the values that best fit to the data, which means the shape of the 0th order function is mostly close to the averaged shape of a set of stellar profiles. In this case, the expansion is made sure to be very compact.

Using these basis functions, we can reproduce the star images as follows:

1) Calculate the center and ellipticity for each stellar profile using the fast fitting algorithms \citealt{2012arXiv1203.0571L}.

2) According to the center and ellipticity of each star, the shape parameters of Gaussian and Moffat model are also calculated using the  fast fitting algorithm (see \citealt{2012arXiv1203.0571L}). \citealt{2001MNRAS.328..977T} argued that a Moffat function could be used to reliably model the turbulence prediction when $\beta\sim4.765$. However,the PSFs usually measured in real images have bigger ¡°wings¡±, or equivalently smaller values of $\beta$ ($2.5<\beta<4$; see \citealt{1993MNRAS.264..961S}), than those expected from the turbulence theory. In this paper, we simply set $\beta=3.5$.

(3) Calculate the mean of the best-fitting parameters over all stars. And use the mean as the value of parameter(s) in the basis functions to create a set of basis functions we will use then.

(4) Finally for each star image, we squeeze the circular symmetric basis functions to the same ellipticity as the star has by performing coordinate transformation and then decompose the star image into several elliptical basis functions. The basis functions have to be pixelated on a finite region in order to perform numerical simulation. This causes violation of the orthogonality and the coefficients of each basis function can not be simply derived from inner product . We overcome this difficulty by solving a Maximum-likelihood solution(e.g., \citealt{2011MNRAS.417.2465A}).

\begin{figure}[!h]
\centering
\includegraphics[scale=0.5]{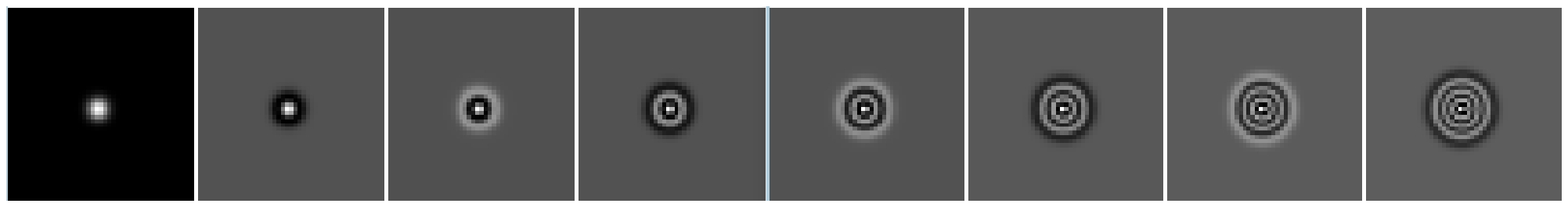}
\includegraphics[scale=0.5]{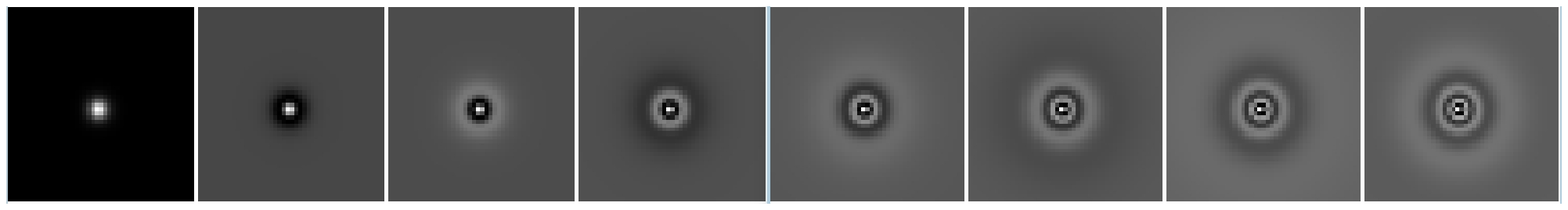}
%\begin{minipage}[]{85mm}
\caption{A demonstration of the first eight basis functions of Gaussianlets (first row) and Moffatlets (second row) constructed in the program respectively. From left to right, as the order of the basis functions increases, the number of ¡°circles¡± increase. As you will also notice, Moffatlets basis functions are "fatter" than Gaussianlets basis functions.}
%\end{minipage}
\label{Fig1}
\end{figure}

Finally we have two parameters ($e_1,e_2$) and several coefficients of basis functions for each star. The first eight basis functions of Gaussianlets (first row) and Moffatlets (second row) are shown in Fig.~\ref{Fig1}. We can see that Moffatlets are more extended than Gaussianlets. This property will lead Moffatlets method to show good performance in the reconstruction of star image.

\subsection{EMPCA}
\label{sect:2.2}
The EMPCA method is an extended version of the classical PCA. It uses expectation and maximization steps to subtract the eigenvectors. The most important improvement is that the noise in the data can also be taken into account (\citealt{2012PASP..124.1015B}). These improvements provide a high-efficiency calculation and reasonable handling of noise. We adopts per-variable weight strategy in this work which can be summarized as follows. The $\chi^2-$fucntion is defined as
\begin{equation}
\chi^2=\sum_{var i, obs j}w_{ij}(A_{ij}-\phi_ic_j)^2,
\label{eq:7}
\end{equation}
For given eigenvector $\phi$, the E-step gives the optimal coefficient as:
\begin{equation}
c_j\longleftarrow\frac{\sum_iw_{ij}A_{ij}}{\sum_iw_{ij}\phi^2_i},
\label{eq:8}
\end{equation}
Then the M-step improves the eigenvector as:
\begin{equation}
\phi_i\longleftarrow\frac{\sum_jw_{ij}c_jA_{ij}}{\sum_jw_{ij}c^2_j}.
\label{eq:9}
\end{equation}
where $A_{ij}=(a_1,\dots,a_{Nstar})$ is initially the dataset in which $a_j$ is a vector denoting the $j$th star, $\phi_i$ is initially the 1st PC we are searching for and $c_j$ is the decomposition coefficient of the $j$th star on the PC. The goal is to solve the minimization problem of (\ref{eq:7})  incorporating a weights matrix $w_{ij}$. The algorithm starts with an arbitrary $\phi$ , and then updates $\phi$ through E-step and M-step iteratively until converged.
To find the higher-order PCs, we replace $A$ by $(A-\phi c)$ and repeat the above process. This procedure can be continued until there are no more effective PCs show up.

The weight is simply related to the noise in each pixels as $w_{ij}=1/\sigma^2_{ij}$. Our simulated stars contain Poisson and Gaussian noise. The estimation of $\sigma_{ij}$ is given by the following rule: a Gaussian noise $\sigma$ is evaluated on the outskirt of each star stamps, then for pixel with value $I$ smaller than $2\sigma$, we take $\sigma_{ij}=\sigma$, for pixel value $I$ larger than $2\sigma$, we take $\sigma_{ij}=\sqrt{\sigma^2+gI}$, where $g$ is the gain of CCD.

\section{Data description}
\label{sect:3}
In this paper, we invoke PhoSim (\citealt{2015ApJS..218...14P}), our primary tool for generating simulated images. PhoSim uses a photon Monte Carlo approach to construct images by sampling photons from models of astronomical source populations. PhoSim is designed to represent Large Synoptic Survey Telescope (LSST) performance and generates images expected for LSST with high fidelity. All detailed atmosphere, telescope and camera physical effects  that determine the shapes, locations and brightnesses of individual stars and galaxies can be accurately represented. This makes PhoSim a perfect simulation tool for study of PSF.

To examine the PSF effects  four images are generated using PhoSim version 3.4.  We simulate images for two LSST chips: R22\_S11 indicates center chip in the focal plane while R02\_R01 indicates a chip near the edge of the focal plane.  In PhoSim all the physical effects can be separately turned 'on' and 'off' so that we can have control over the effects which may affect PSF.  Two of the images are simulated with diffraction 'off' and the other two with diffraction 'on'.   For all the other physical effects default settings of LSST are used.

In the four images, the pixel size is $0.2^{''}$/pixel and the pixel values are simulated in ADU unit with gain=1, hence the value on each pixel counts the number of photons fall in. The star images contains only Poisson noise and all have roughly the same magnitude. We can then add different amounts of background Gaussian noise to each star and estimate the signal-to-noise ratio (SNR) within a $10\times10$ square around the brightest pixel by using definition SNR$=\Sigma_p(I_p)/\sqrt{\Sigma_p(\sigma^2+I_p)}$, where $I_p$ is the LSST simulator data with only Poisson noise, $\sigma$ is the Gaussian noise we added in later.

Based on our simulation, an explicit model fitting is also performed as a verification of the result in \citealt{2001MNRAS.328..977T}, they claimed that the Moffat function is the better model for fitting PSFs than Gaussian function. We choose several bright stars from our simulation and then fit them with these two models respectively by minimizing the $\chi^2$ function:
\begin{equation}
\chi^2=\sum_{i, j}\frac{(I_{ij}^0-F_{ij})^2}{\sigma^2_{ij}},
\label{eq:12}
\end{equation}
where $F_{ij}$ is the fitting function and $\sigma_{ij}$ is the Possion noise in each pixel.

\begin{figure}[!h]
\centering
\includegraphics[scale=0.7]{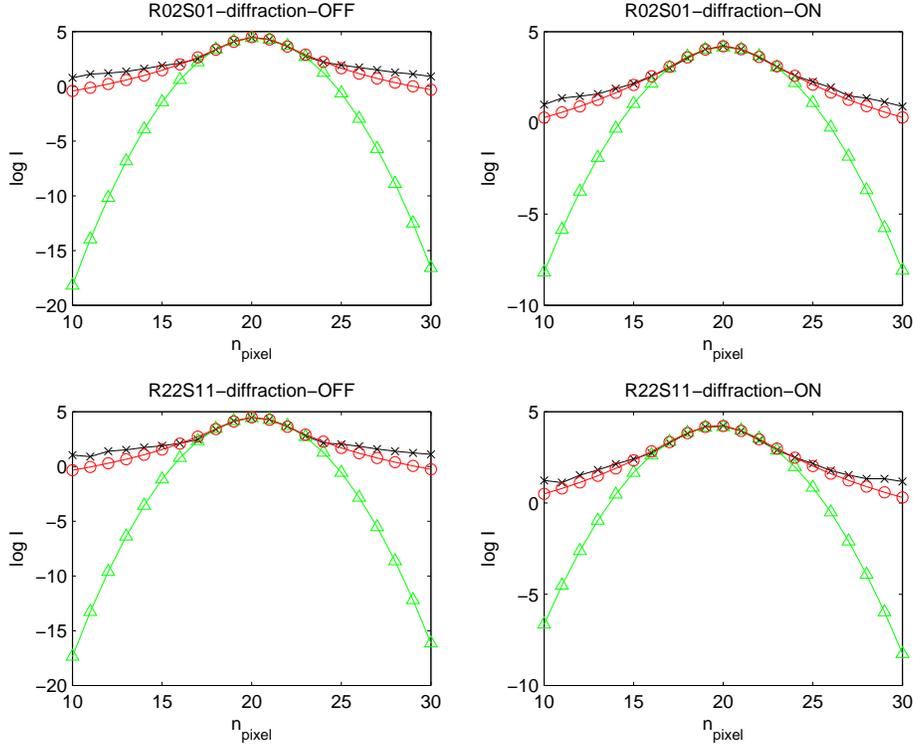}
%\begin{minipage}[]{85mm}
\caption{After computing the best fits for parameters $\sigma$ and $r_d$, we construct the corresponding Gaussian (green dots and lines) and Moffat (red dots and lines) functions and compare to the original star profiles (black dots and lines). The drawn curves are only the cross-sections of the 2-dim images.}
%\end{minipage}
\label{Fig2}
\end{figure}

Fig.\ref{Fig2} shows the fitted brightness profile along the diagonal for our four kinds of simulated stars. Clearly, Moffat model can fit our simulated stars much better than Gaussian profile. The Gaussian profile only fits the inner part well but drops too fast in the outer region. Meanwhile the fitting of the Moffat model is well behaved even at large radius. Therefore we would expect that Moffatlets will also work better than Gaussianlets in the following tests.

\begin{figure}[!h]
%begin{tabular}{cc}
\begin{minipage}{0.48\linewidth}
\centerline{\includegraphics[width=6.5cm]{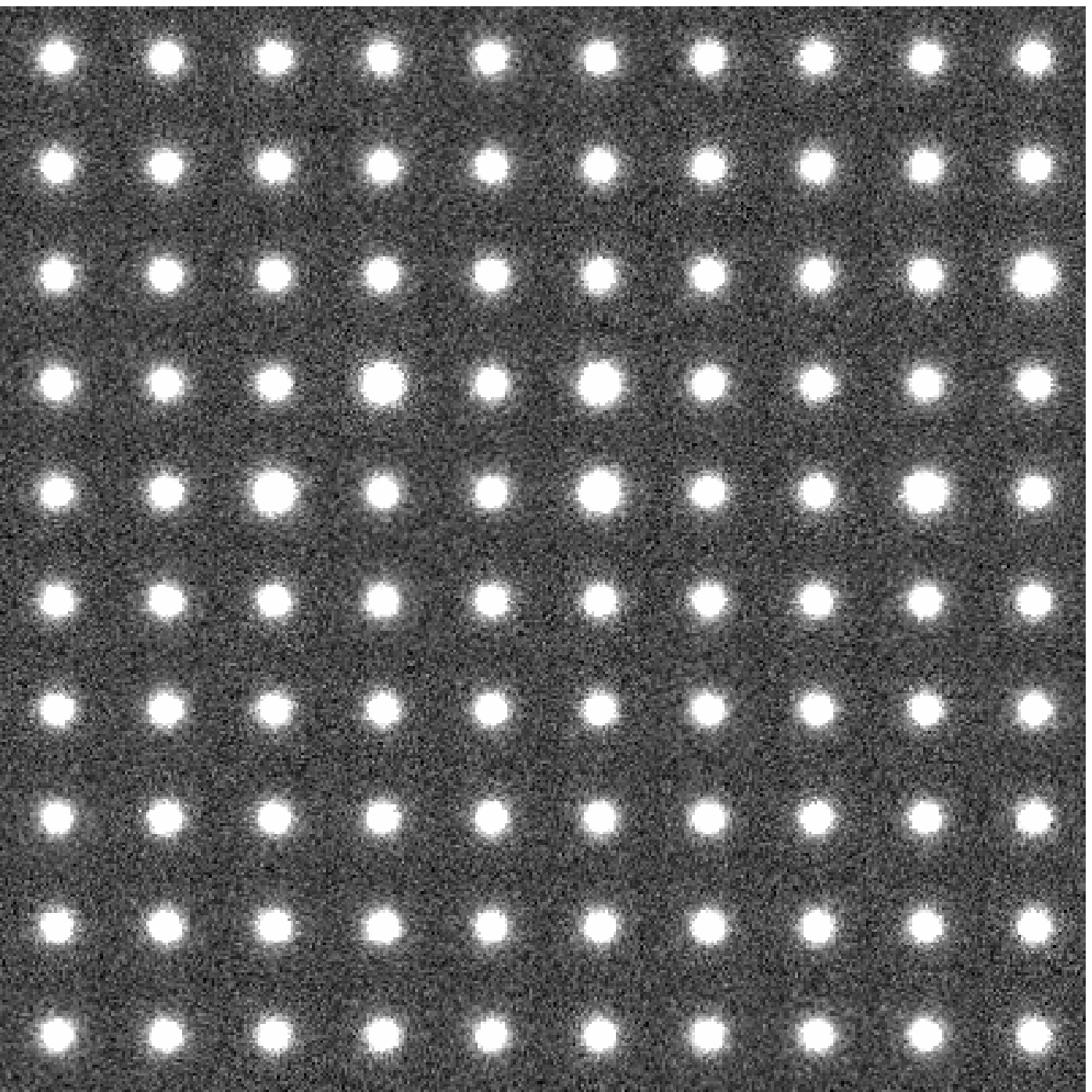}}
\centerline{(a) $\sigma=10$}
\end{minipage}
\hfill
\begin{minipage}{0.48\linewidth}
\centerline{\includegraphics[width=6.5cm]{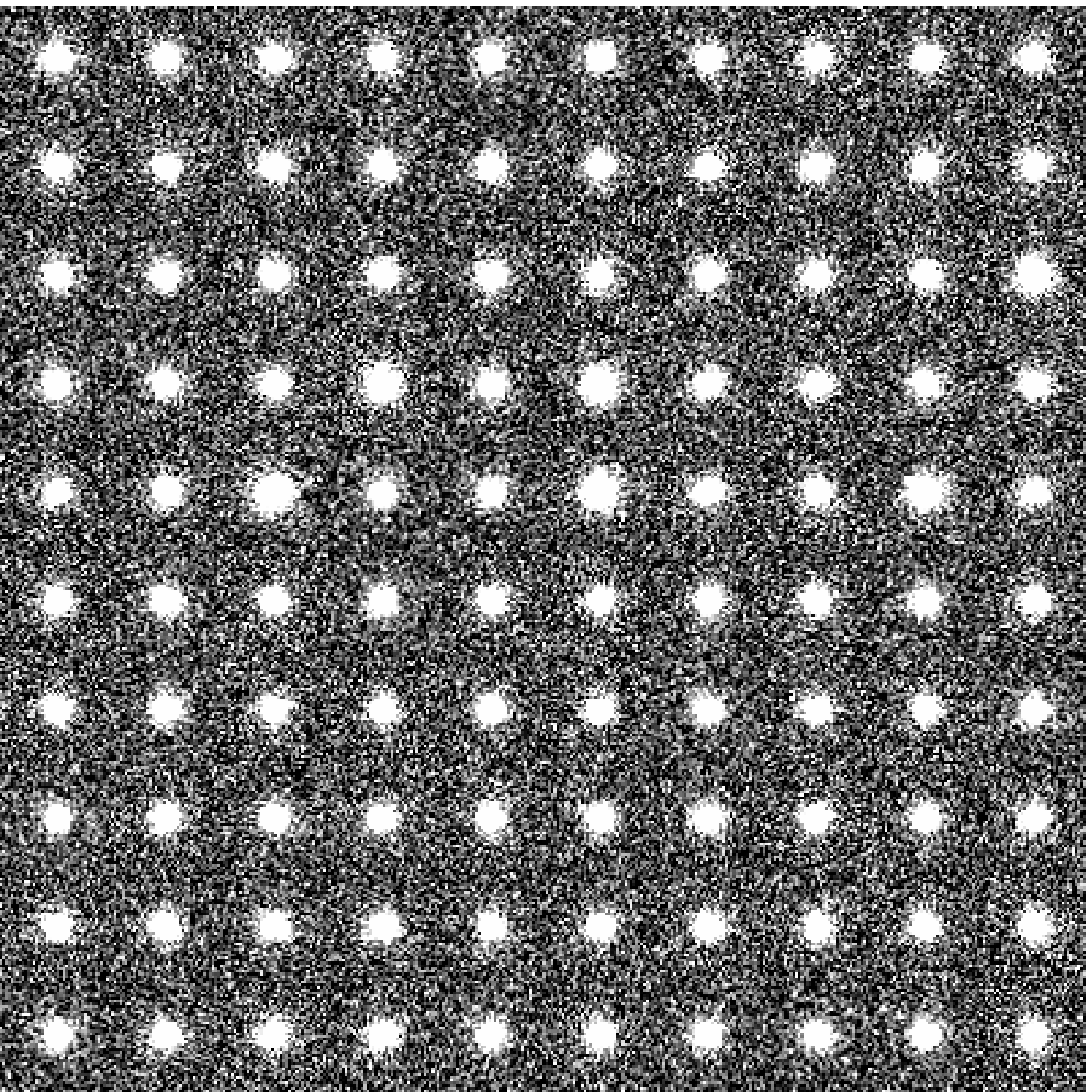}}
\centerline{(b) $\sigma=40$}
\end{minipage}
\vfill
\begin{minipage}{0.48\linewidth}
\centerline{\includegraphics[width=6.5cm]{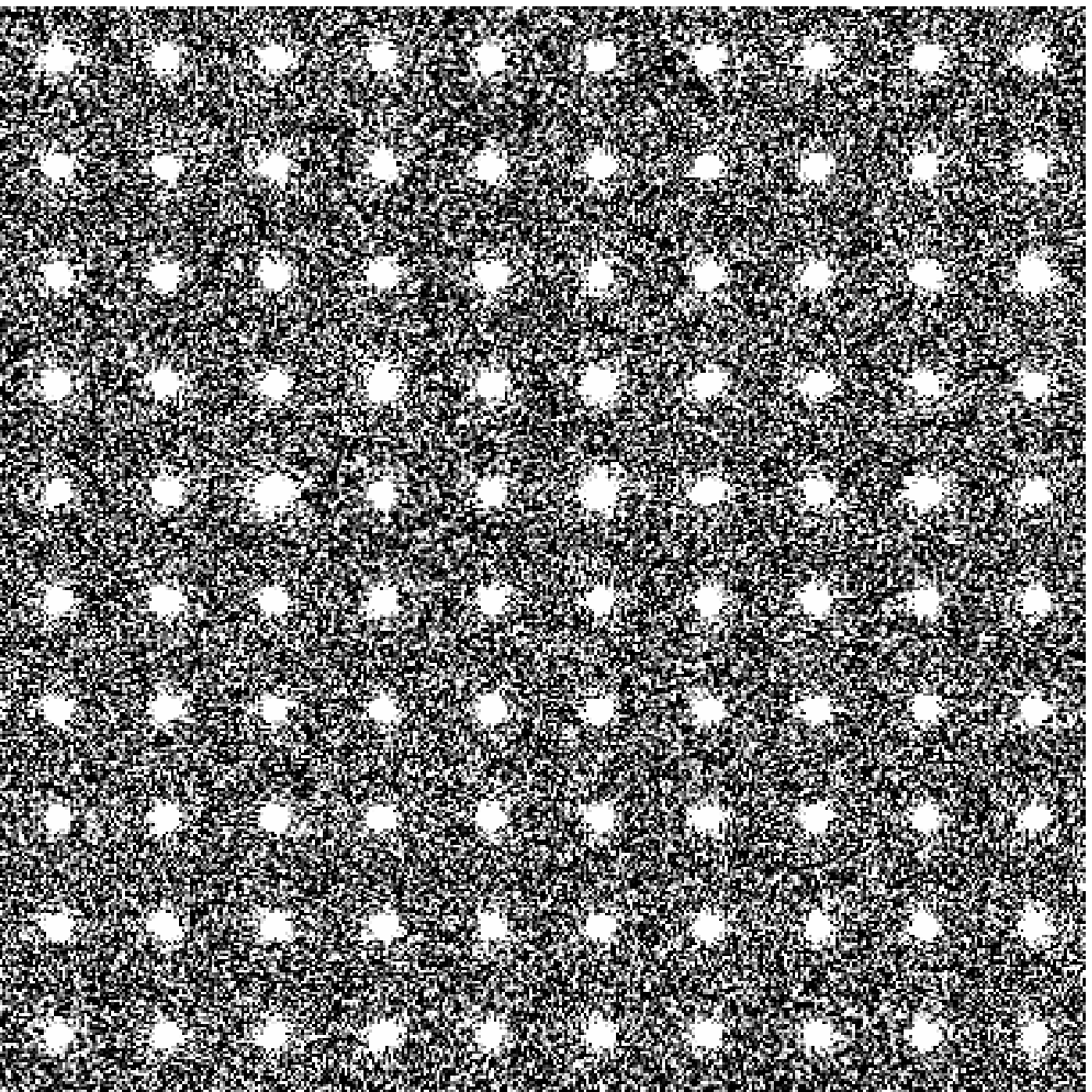}}
\centerline{(c) $\sigma=80$}
\end{minipage}
\hfill
\begin{minipage}{0.48\linewidth}
\centerline{\includegraphics[width=6.5cm]{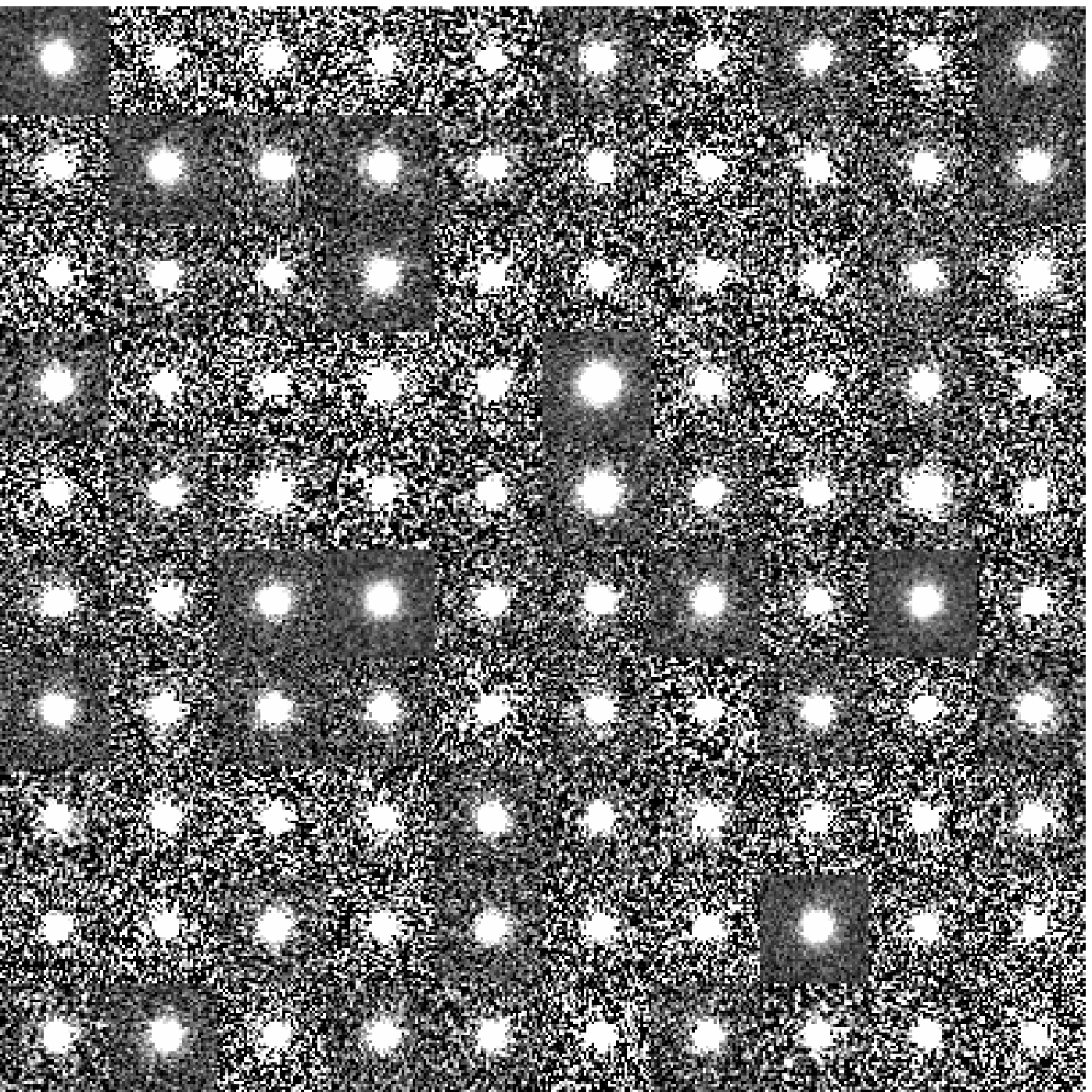}}
\centerline{(d) $\sigma=ran(10-100)$}
\end{minipage}
%end{tabular}
\caption{The four panels show background noise added in four different cases. All pictures are drawn in the same gray scale.}
\label{Fig3}
\end{figure}

One hundred stars are randomly selected from each CCD. The test is then divided into four cases: In the first three cases, an uniform level of Gaussian noise is added to the 100 stars with $\sigma=10,40,80$ (with $\langle$SNR$\rangle$$\approx416, 232$ and $132$) respectively; in the fourth case, different amounts of Gaussian noise with the value of $\sigma$ randomly ranging from 10 to 100 are added to the stars. By including the last case, we try to mimic 100 stars of different SNR ($80\lesssim$SNR$\lesssim500$), or in terms of observations, there are 100 stars with different magnitudes.Fig.\ref{Fig3} shows these four cases of R02S01-diffraction-OFF simulation.

\section{Results}           %% first-level sections will be auto-capitalized

The reconstructions using the three methods are performed on the noise interfered data in all four cases. we aim to test how efficient are these three methods and how the noise affect the results. For the basis function methods, the shape parameters are computed according to the average shapes of the 100 stars. The value of $\sigma_d$ and $r_d$ in the Gaussianlets and Moffatltes are listed in Table.~\ref{Table1} and Table.~\ref{Table2} respectively. It shows that value of $\sigma_d$ and $r_d$ taken for "diffraction on" data are larger than "diffraction off" data. This is because spikes exist in "diffraction on" data and make the stars more extended. Once the parameters are fitted, we can create the basis functions. Fig.~\ref{Fig1} demonstrates the first 8 basis functions of Gaussianlets and Moffatlets respectively.

\label{sect:4}
\begin{table}[!hbp]
\begin{tabular}{|c|c|c|c|c|}
\hline
  & 10 & 40 & 80 & ran(10-100) \\
\hline
R02S01-diffraction-OFF & 1.419150 & 1.419232 & 1.419417 & 1.419747 \\
\hline
R02S01-diffraction-ON & 1.906670 & 1.904350 & 1.904945 & 1.906408 \\
\hline
R22S11-diffraction-OFF & 1.443192 & 1.442398 & 1.442060 & 1.440537 \\
\hline
R22S11-diffraction-ON & 1.921579 & 1.919953 & 1.920336 & 1.921725 \\
\hline
\end{tabular}
\caption{Value of $\sigma_d$ taken in different simulation runs}
\label{Table1}
\end{table}

\begin{table}[!hbp]
\begin{tabular}{|c|c|c|c|c|}
\hline
  & 10 & 40 & 80 & ran(10-100) \\
\hline
R02S01-diffraction-OFF & 3.030692 & 3.029967 & 3.029196 & 3.029439 \\
\hline
R02S01-diffraction-ON & 4.061016 & 4.052640 & 4.051725 & 4.054829 \\
\hline
R22S11-diffraction-OFF & 3.083501 & 3.080003 & 3.077113 & 3.072700 \\
\hline
R22S11-diffraction-ON & 4.094765 & 4.089887 & 4.085220 & 4.086687 \\
\hline
\end{tabular}
\caption{Value of $r_d$ taken in different simulation runs}
\label{Table2}
\end{table}

\begin{figure}[!h]
\centering
\includegraphics[scale=0.5]{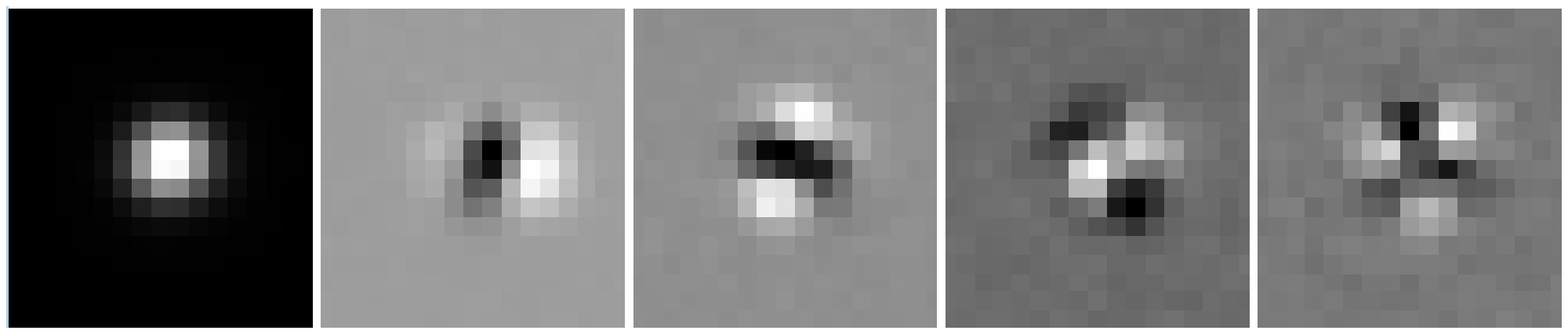}
\includegraphics[scale=0.5]{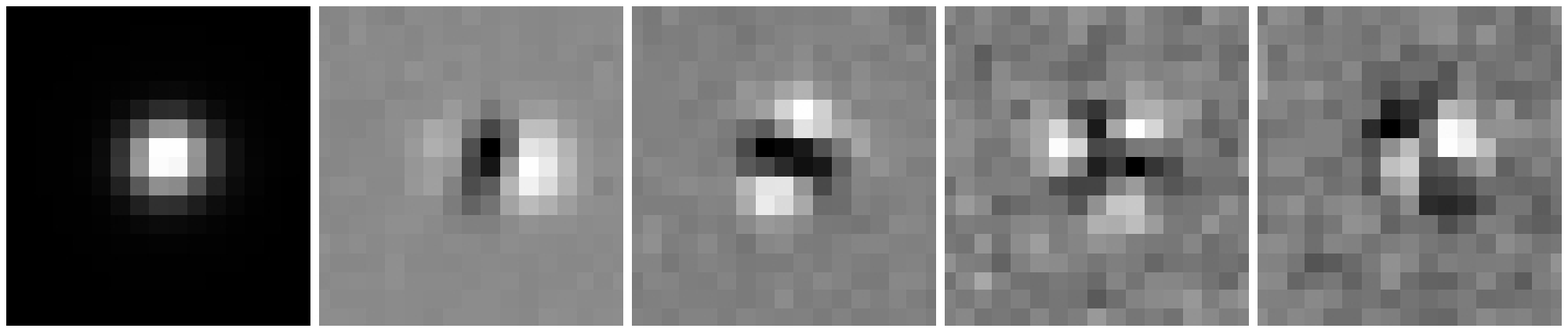}
\includegraphics[scale=0.5]{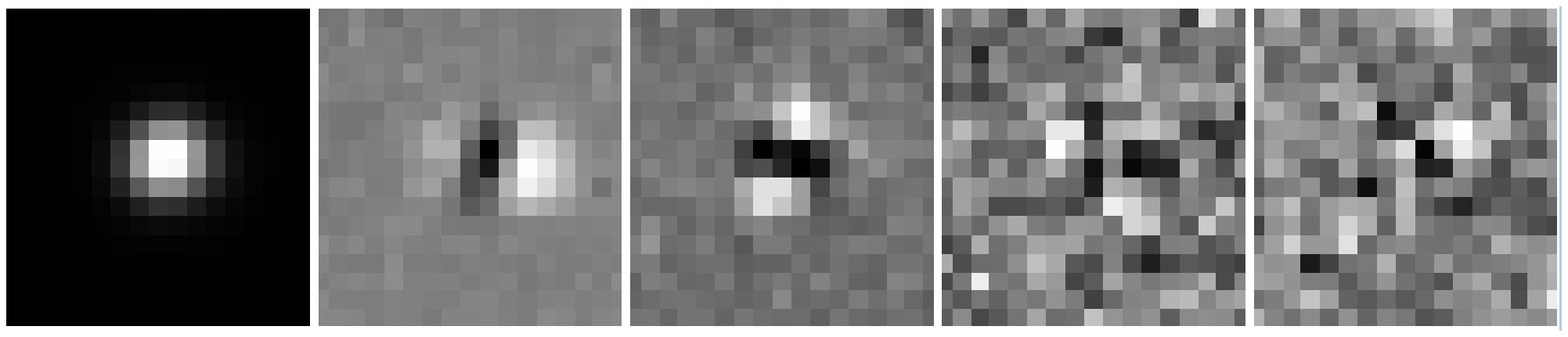}
\includegraphics[scale=0.5]{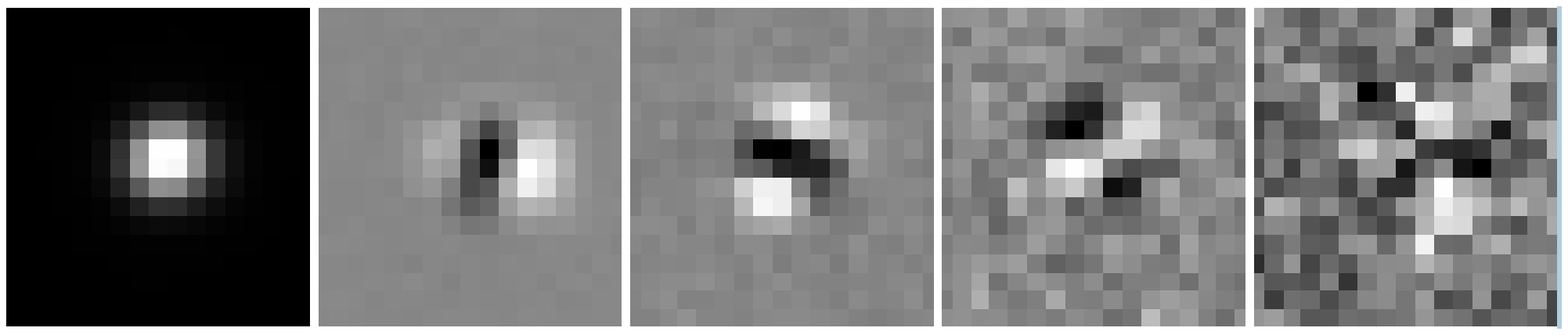}
%\begin{minipage}[]{85mm}
\caption{The first five PCs extracted in different cases. The first row is the case that we add to the data Gauss noise of $\sigma=10$. In this case we can see the most amount of signals. The other three rows are cases of $\sigma=40,80$ and $ran(10-100)$ respectively. As the noise increased, less and less signal can be recognized in the higher order PCs.}
%\end{minipage}
\label{Fig4}
\end{figure}

Noise affects the EMPCA method in a very apparent way. In Fig.~\ref{Fig4}, a set of patterns were clearly resolved by the PCA algorithm in case 1($\sigma=10$). But as more noise is added, less useful PCs will be extracted. Just as shown in case 2($\sigma=40$), case 3($\sigma=80$) and case 4($\sigma=ran(10-100)$), all the high-order PCs contain visible noise and no signal can even be easily recognized in some of them. This introduces uncertainty to appropriately choose the number of the PCs to be used in the EMPCA method.

Here we introduce the usual $\chi^2$ function to quantify how well the reconstruction is done.
\begin{equation}
\chi^2=\sum_{i, j}\frac{(I_{ij}^0-I_{ij}^{(reconstructed)})^2w_{ij}}{N_{pixels}},
\label{eq:10}
\end{equation}
$I^0$ refers to the original extracted star with poisson noise only. The weight is the same as we employed in the EMPCA algorithm.

\begin{figure}[!h]
\centering
\includegraphics[scale=0.6]{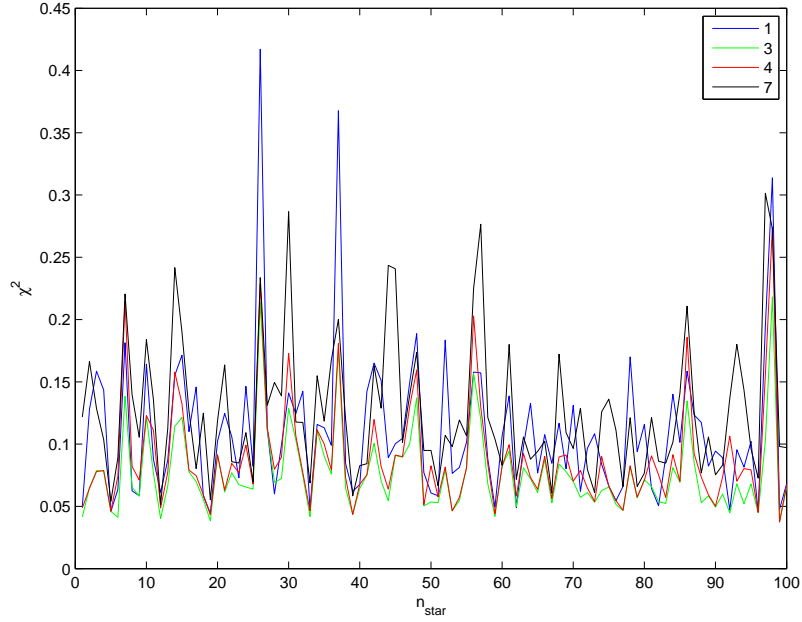}
%\begin{minipage}[]{85mm}
\caption{The $\chi^2$ of reconstructions by using different number of PCs are compared. $n_{star}$ is the id of the star. The blue line correspond to reconstructions done by only 1 PC. The green, red and black lines are that of 3,4 and 7 PCs used correspondingly. The green line is the lowest one indicating the best fit. }
%\end{minipage}
\label{Fig5}
\end{figure}

We draw the $\chi^2-$curves corresponding to reconstructions using different number of PCs (Fig.~\ref{Fig5}) (The reconstruction is performed on image R02S01-diffraction-OFF,$\sigma=40$). The green line which corresponds to using 3 PCs is the lowest one. As we decrease the number of used PCs, the $\chi^2-$curve raises up (as the black line of 1 PC indicates) since fewer PCs means less information is taken into account. As we increase the number of used PCs, the $\chi^2-$curve also raises up due to the fact that more noise are contained in the reconstructed star images. In fact, we can see that using 7 PCs is worse than using just 1 PC. In the following reconstructions, applied to all data sets, we always use 4 PCs.

As for our basis function methods, we would expect more number of free parameters because the theoretical basis functions can not be more compact than the numerically solved PCs. Since there are already two parameters($e_1,e_1$) for each star, we simply adopt four basis functions for Moffatlets and Gaussianlets method.

\begin{figure}[!h]
\centering
\includegraphics[scale=0.7]{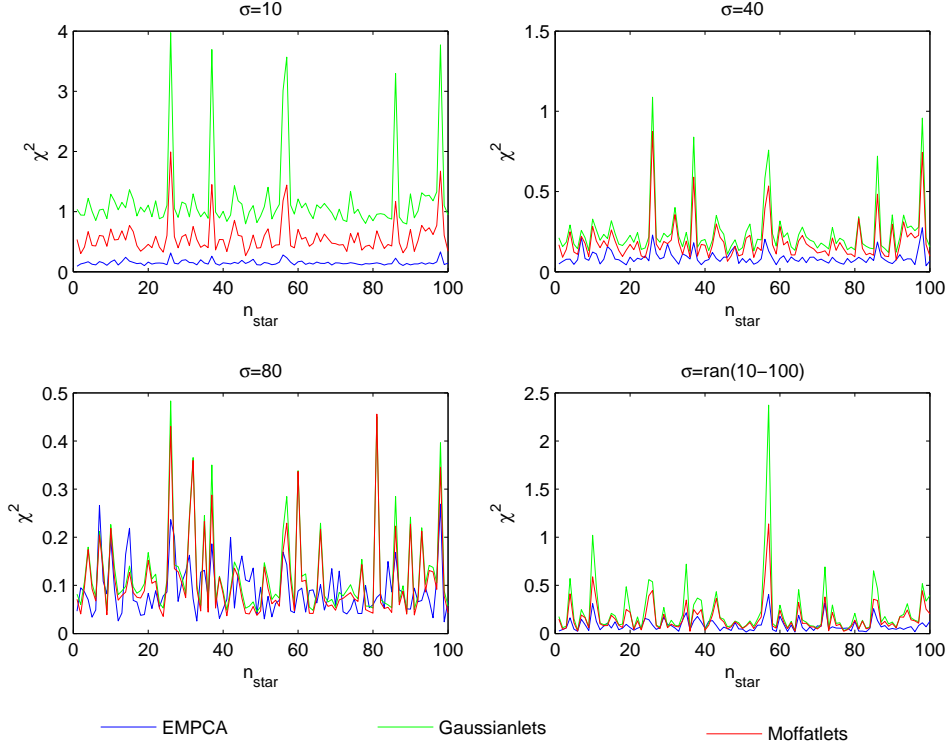}
%\begin{minipage}[]{85mm}
\caption{$\chi^2$-curves in the test of data set R02S01-diffraction-OFF. Blue lines are for EMPCA, green lines for Gaussianlets and red lines for Moffatlets. The same denotement is adopted in the next $\chi^2-$plots Fig.~\ref{Fig7},~\ref{Fig8},~\ref{Fig9} and $\delta R^2-$plots Fig.~\ref{Fig10},~\ref{Fig11},~\ref{Fig12},~\ref{Fig13}.}
%\end{minipage}
\label{Fig6}
\end{figure}

\begin{figure}[!h]
\centering
\includegraphics[scale=0.7]{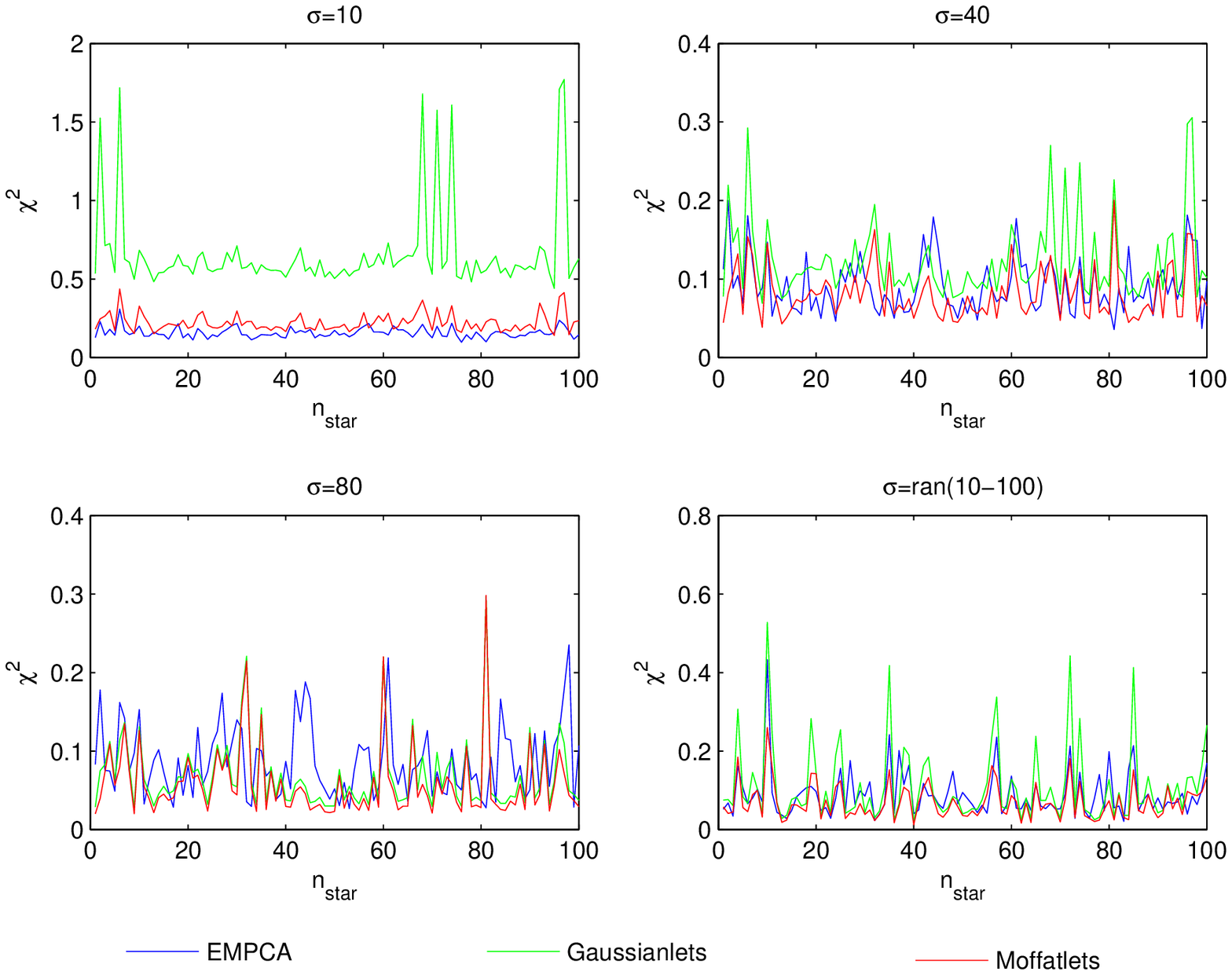}
%\begin{minipage}[]{85mm}
\caption{$\chi^2$-curves in the test of data set R02S01-diffraction-ON.}
%\end{minipage}
\label{Fig7}
\end{figure}

\begin{figure}[!h]
\centering
\includegraphics[scale=0.7]{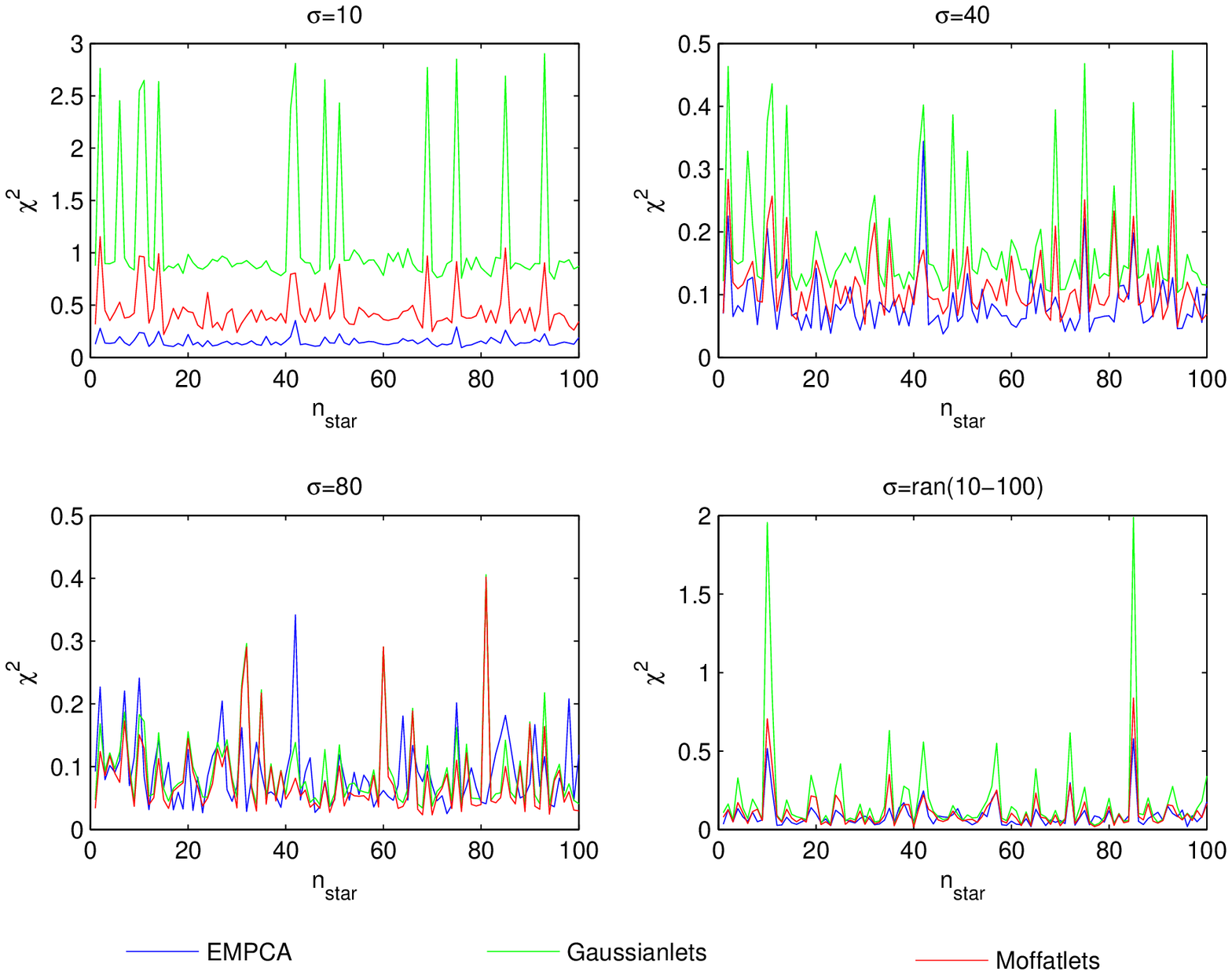}
%\begin{minipage}[]{85mm}
\caption{$\chi^2$-curves in the test of data set R22S11-diffraction-OFF.}
%\end{minipage}
\label{Fig8}
\end{figure}

\begin{figure}[!h]
\centering
\includegraphics[scale=0.7]{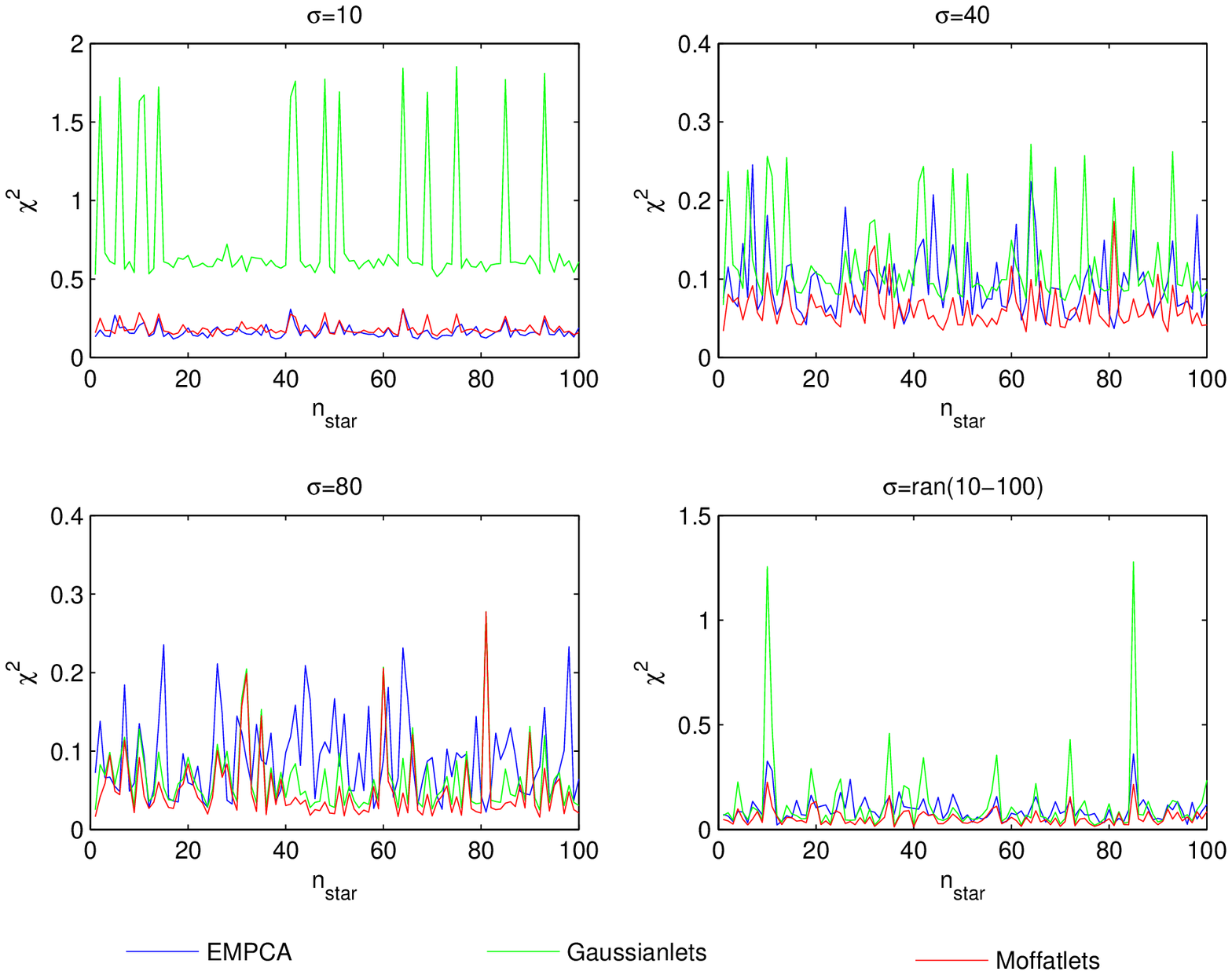}
%\begin{minipage}[]{85mm}
\caption{$\chi^2$-curves in the test of data set R22S11-diffraction-ON.}
%\end{minipage}
\label{Fig9}
\end{figure}

Using the $\chi^2$ quantity, the three methods are compared for all data sets in Fig.~\ref{Fig6},~\ref{Fig7},~\ref{Fig8},~\ref{Fig9}, where the blue, red and green lines are for the results of EMPCA ,Moffatlets and Gaussianlets respectively. All of these results broadly support the same conclusions:1) EMPCA always performs better than Moffatlets and Gaussianlets for the high SNR cases. This is because there are high-order patterns of brightness distribution in star images which can not be described by our elliptical basis functions but can be resolved by EMPCA. But for the low SNR cases, the results of EMPCA are comparable with the results of Moffatlets.
2)Moffatlets is always better than Gaussianlets. As mentioned in the introduction and also in \citealt{2001MNRAS.328..977T}, that Moffat function fit real PSF better than Gaussian function. Gaussianlets performs especially poorly in case 1 since the Gaussian function cannot describe the presence of large "wings" in PSF which is not buried by noise in this case. The several high peaks indicated by green line in case 4($sigma=ran(10-100)$) correspond the stars with higher SNRs.
We also see the "diffraction on" stars can be reconstructed better than "diffraction off" stars. This is because in our simulation the diffraction spikes are not very sharp rather it makes the stars more extended and smooth some high order minor substructures.

Another two quantities are also introduced to serve as test for the efficiency of the reconstructions.
The first quantity is the ellipticity defined as:
\begin{equation}
   e_1=\frac{Q_{11}-Q_{22}}{Q_{11}+Q_{22}},
\label{eq:5}
\end{equation}
\begin{equation}
   e_2=\frac{2Q_{12}}{Q_{11}+Q_{22}}.
\label{eq:6}
\end{equation}
where $Q{ij}$ are second brightness moments of star image.  A Gaussian filter is employed in calculating the moments.The FWHM of the Gaussian filter is the mean FWHM of stars.

The other quantity is the square rms size of star defined as,
\begin{equation}
   R^2=Q_{11}^2+Q_{22}^2
\label{eq:12}
\end{equation}

\begin{figure}[!h]
\centering
\includegraphics[scale=0.7]{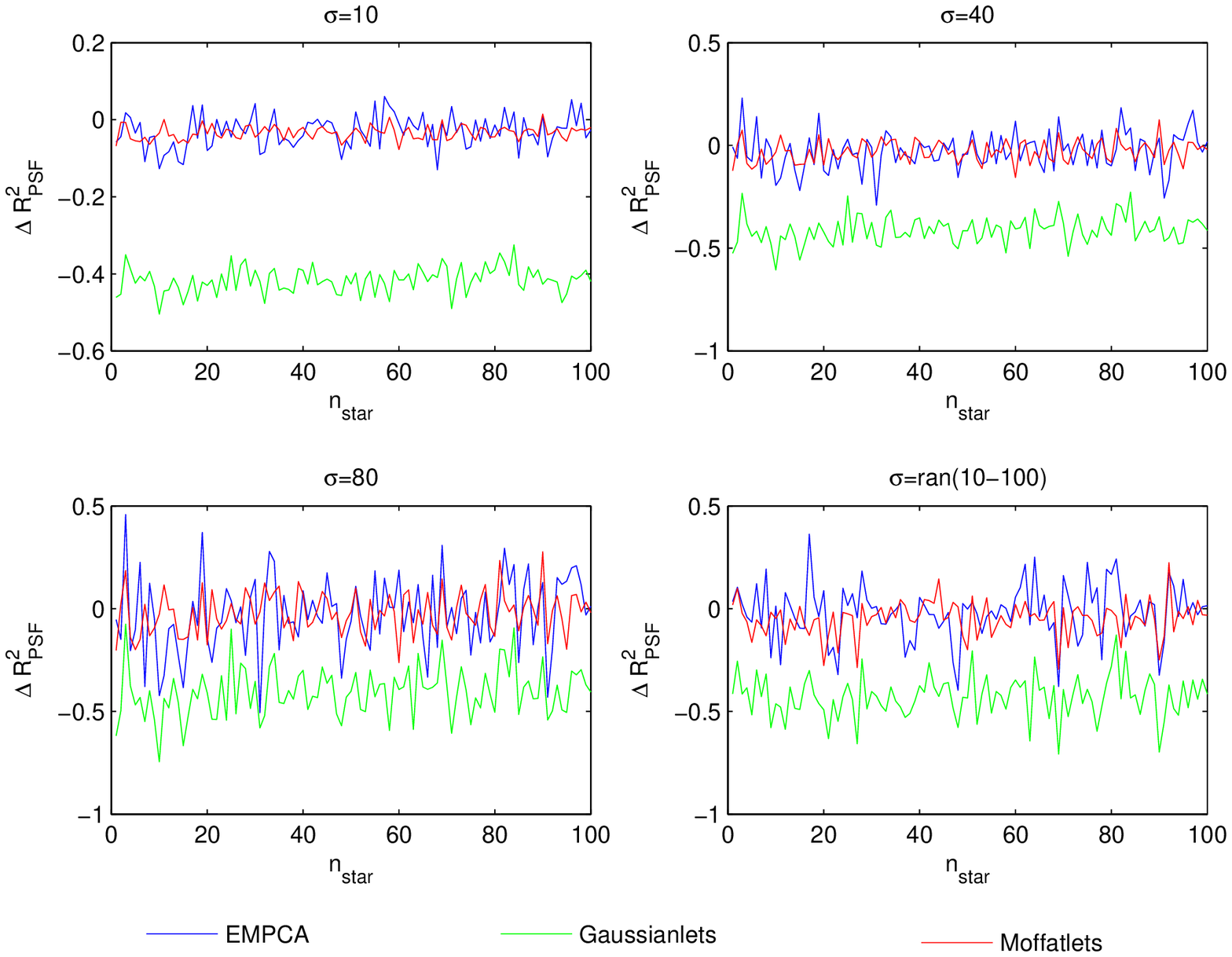}
%\begin{minipage}[]{85mm}
\caption{$\delta R^2$-curves in the test of data set R02S01-diffraction-OFF.}
%\end{minipage}
\label{Fig10}
\end{figure}

\begin{figure}[!h]
\centering
\includegraphics[scale=0.7]{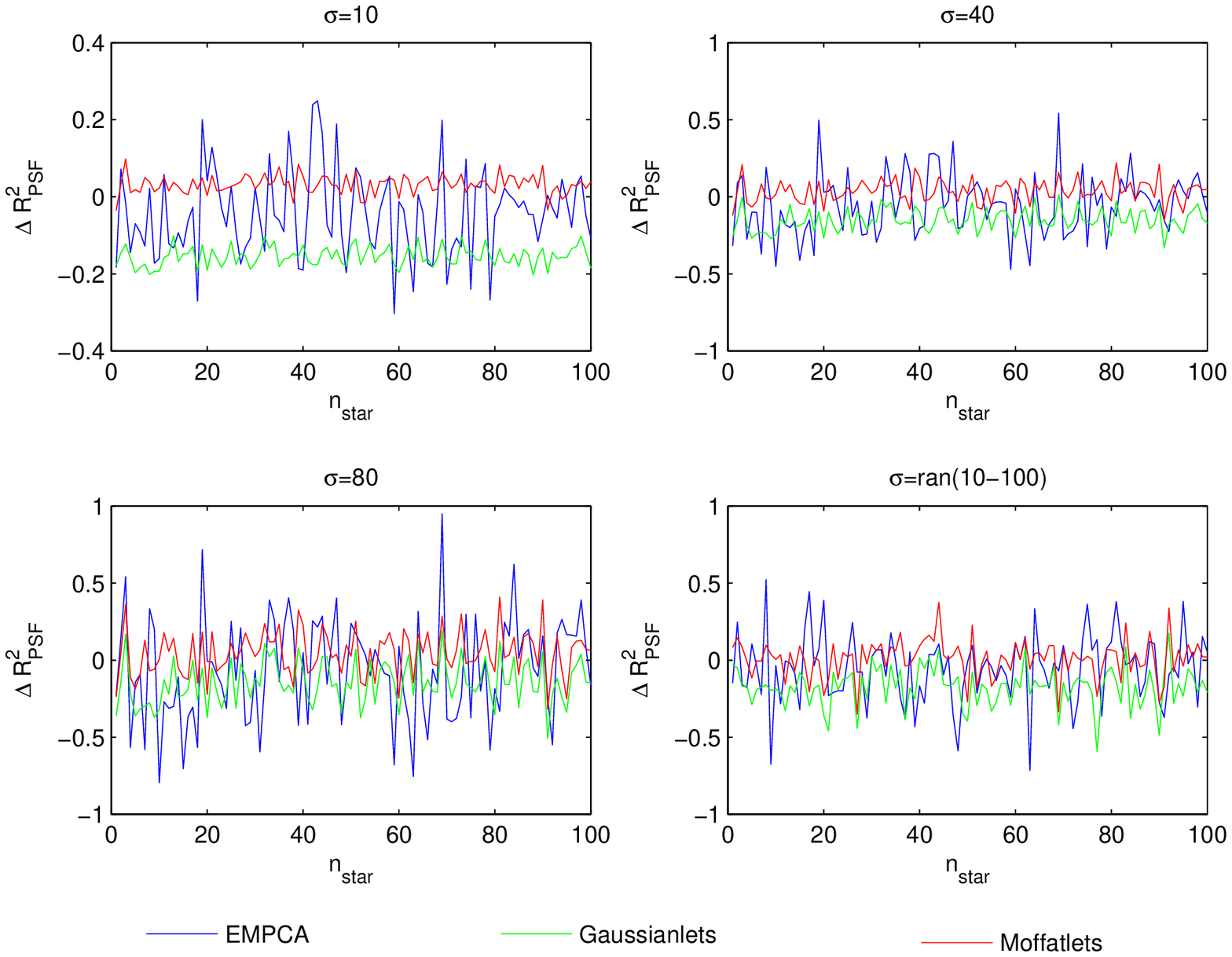}
%\begin{minipage}[]{85mm}
\caption{$\delta R^2$-curves in the test of data set R02S01-diffraction-ON.}
%\end{minipage}
\label{Fig11}
\end{figure}

\begin{figure}[!h]
\centering
\includegraphics[scale=0.7]{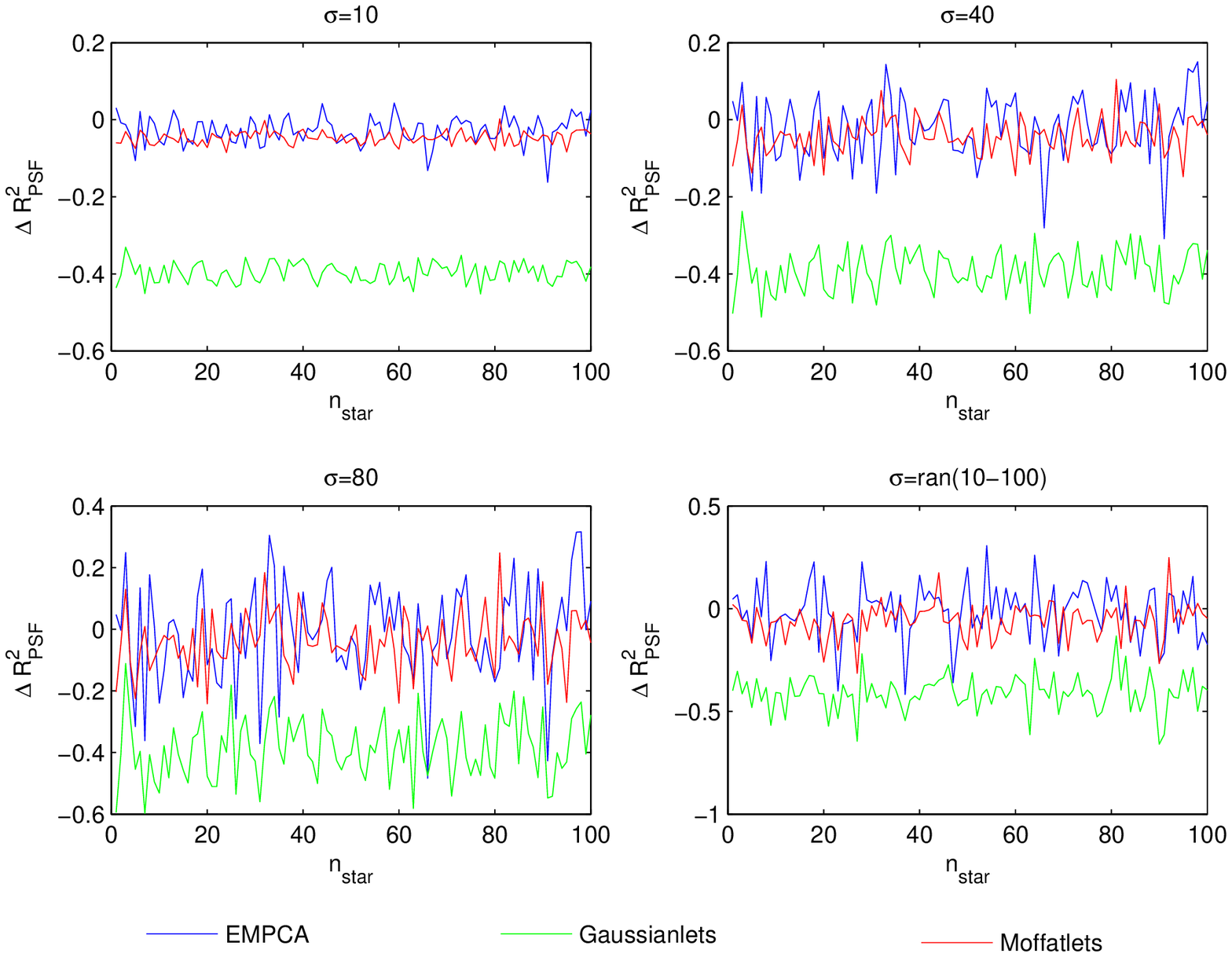}
%\begin{minipage}[]{85mm}
\caption{$\delta R^2$-curves in the test of data set R22S11-diffraction-OFF.}
%\end{minipage}
\label{Fig12}
\end{figure}

\begin{figure}[!h]
\centering
\includegraphics[scale=0.7]{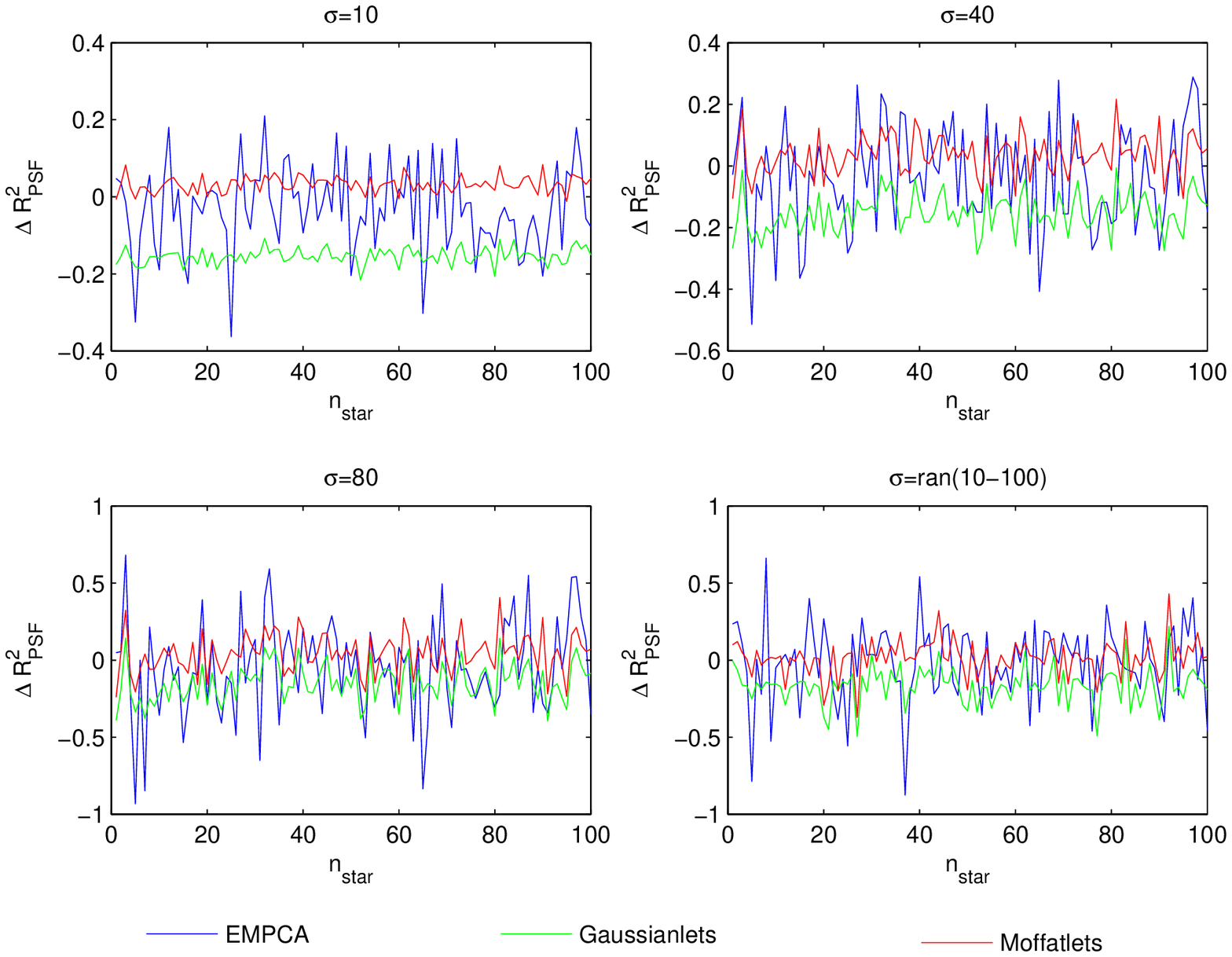}
%\begin{minipage}[]{85mm}
\caption{$\delta R^2$-curves in the test of data set R22S11-diffraction-ON.}
%\end{minipage}
\label{Fig13}
\end{figure}

Using these formulas, we first measure $(R^2,e_1,e_2)$ for the original stars without Gaussian noise added yet. Then we calculate $(R^2,e_1,e_2)$ for the reconstructed stars in each realizations. Finally we compare the differences between the two, $(\delta R^2,\delta e_1,\delta e_2)$. Fig.~\ref{Fig10},~\ref{Fig11},~\ref{Fig12},~\ref{Fig13} compare $\delta R^2$. As above, the blue, red and green lines are for the results of EMPCA, Moffatlets and Gaussianlets respectively. It shows Moffatlets fit the size of stars best (the average $\delta R^2$  over 100 stars is close to 0 and the scatter is very small) and Gaussianlets did worst (the average $\delta R^2$ always biases from 0 a lot although its scatter is also small. As shown in Fig.~\ref{Fig2}, Gaussian function descends too fast at large radius, hence Gaussianlets underestimate the sizes of PSFs with presence of large "wings"). EMPCA also does well but introduces larger scatters because of the noise in the PCs.

\begin{figure}[!h]
\centering
\includegraphics[scale=0.6]{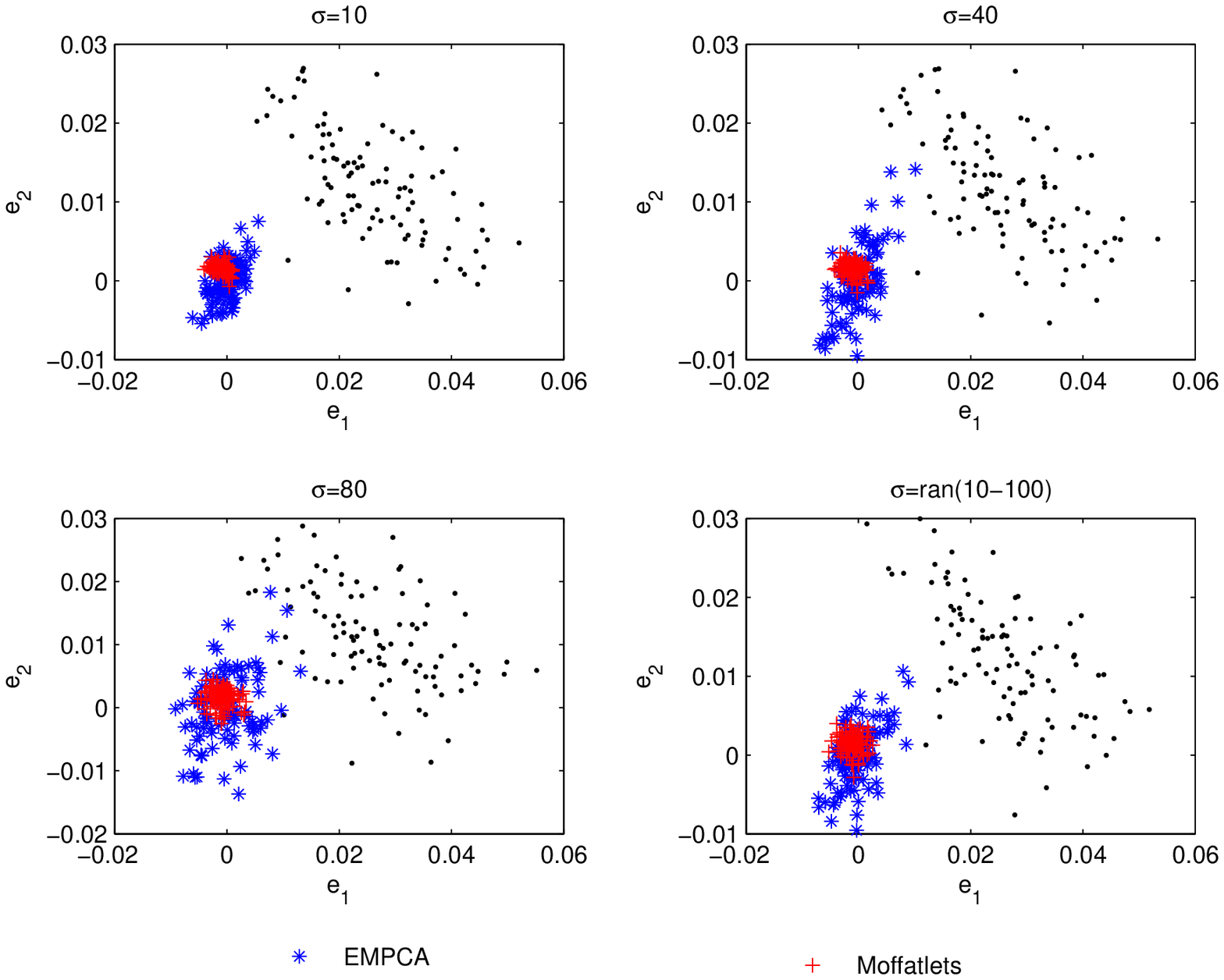}
%\begin{minipage}[]{85mm}
\caption{$(\delta e_1,\delta e_2)$ plotted for the test of data set R02S01-diffraction-OFF. In the figure, there are two kinds of spots: Black spots are ellipticity $(e_1,e_2)$ measured from original stars when background noise are not added yet; Colorful spots (blue and red) are ellipticity difference $(\delta e_1,\delta e_2)$ between the original stars and reconstructed stars using EMPCA and Moffatlets correspondingly. In Fig.~\ref{Fig15},~\ref{Fig16},~\ref{Fig17} we use the same color denotation.}
%\end{minipage}
\label{Fig14}
\end{figure}

\begin{figure}[!h]
\centering
\includegraphics[scale=0.6]{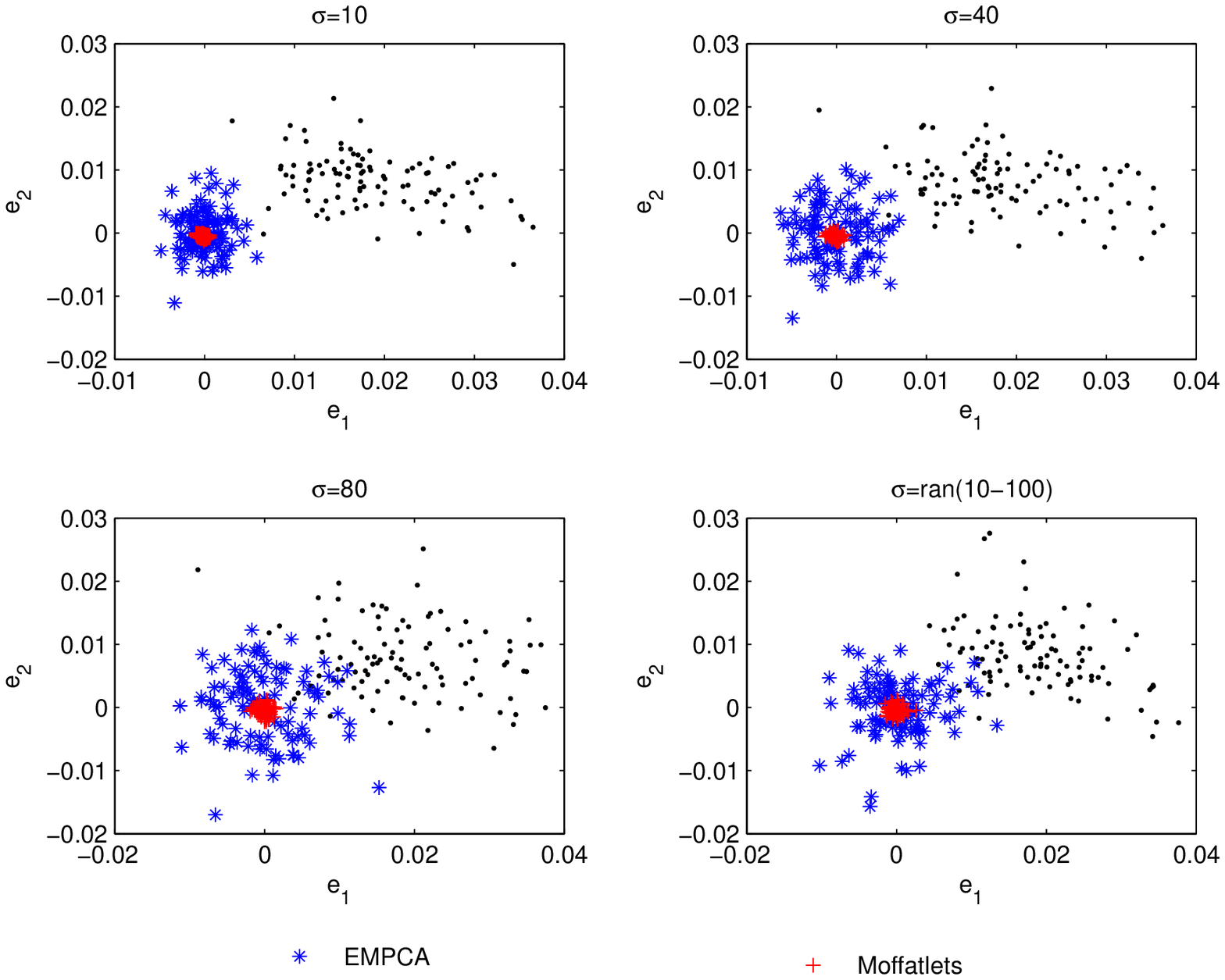}
%\begin{minipage}[]{85mm}
\caption{$(\delta e_1,\delta e_2)$ plotted for the test of data set R02S01-diffraction-ON.}
%\end{minipage}
\label{Fig15}
\end{figure}

\begin{figure}[!h]
\centering
\includegraphics[scale=0.6]{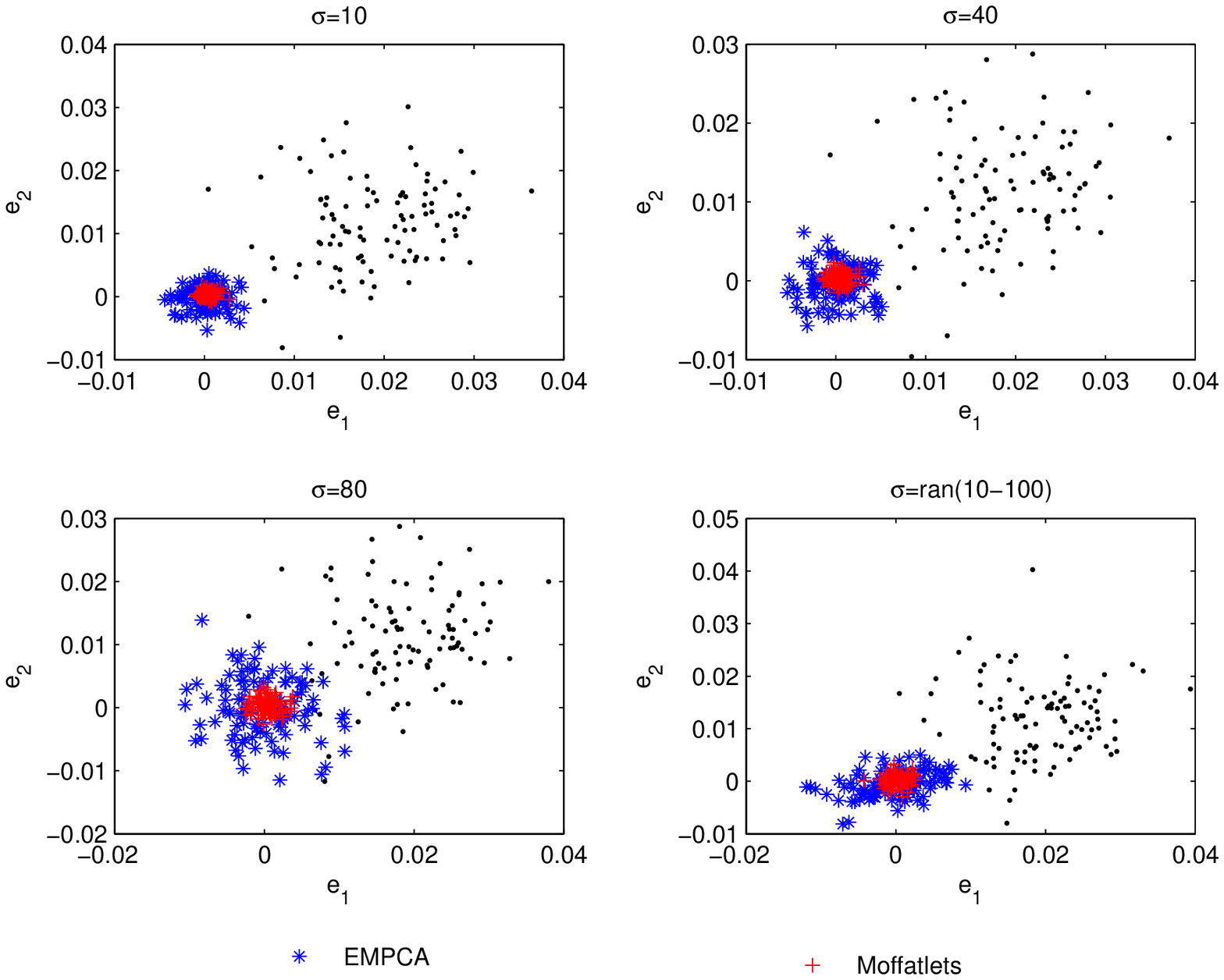}
%\begin{minipage}[]{85mm}
\caption{$(\delta e_1,\delta e_2)$ plotted for the test of data set R22S11-diffraction-OFF.}
%\end{minipage}
\label{Fig16}
\end{figure}

\begin{figure}[!h]
\centering
\includegraphics[scale=0.6]{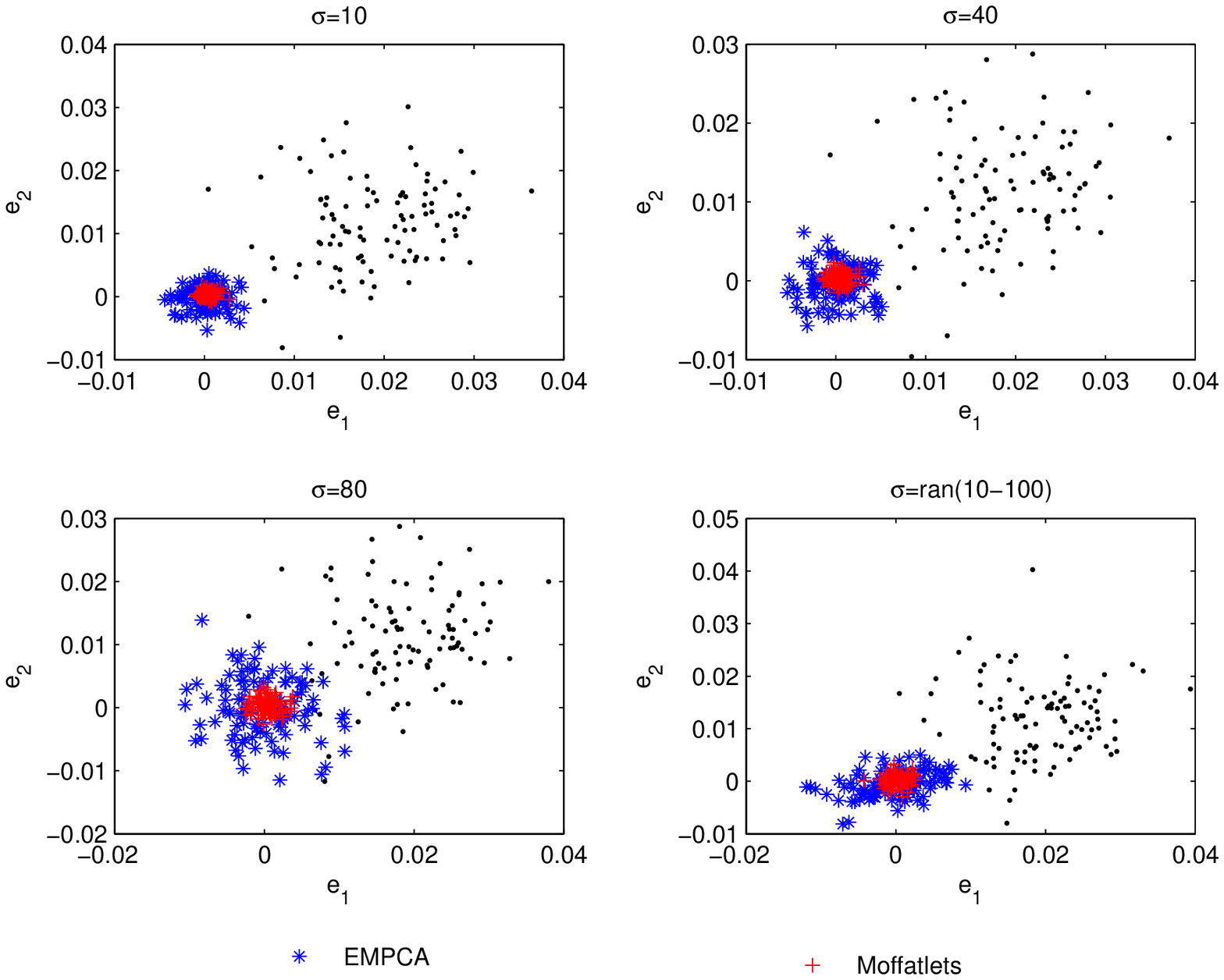}
%\begin{minipage}[]{85mm}
\caption{$(\delta e_1,\delta e_2)$ plotted for the test of data set R22S11-diffraction-ON.}
%\end{minipage}
\label{Fig17}
\end{figure}

Fig.~\ref{Fig14},~\ref{Fig15},~\ref{Fig16},~\ref{Fig17} compare the uncertainty in ellipticity. The black dots show the ellipticity $(e_1,e_2)$ meaured from original stars (without background noise added yet) and the colored dots show the deviations of ellipticity ($\delta e_1, \delta e_2$) between reconstructed stars and original stars. For the basis function methods, the ellipticity are measured first for all stars and the basis functions are then correspondingly reshaped. Therefore the Moffatlets and Gaussianlets method share the same ellipticity and we just plot the results of Moffatlets here. The ellipticity are measured from the Gaussian noise added data, so when the noise increases, the scatter in ellipticity increase.

As we can see that the basis function methods fit the ellipticity of stars much better than EMPCA. Although EMPCA performs well on the $\chi^2$ test, but introduced larger scatters in the size, especially in the ellipticity of PSF. This implies more stars are needed in EMPCA method in order to reconstruct an unknown PSF's size and ellipticity to required accuracy. We calculate $\sigma_{R^2}/R^2$ and $\sigma_e$ in different simulation runs, the results are listed in Table.\ref{Table3},\ref{Table4},\ref{Table5} and Table.\ref{Table6},\ref{Table7}. As shown in \citealt{2008A&A...484...67P}, uncertainties on size and ellipticity of PSF calibration is the two key parameters which will propagate into the systematics of shear measurement. The requirement for systematic bias of the cosmic shear measurement, $\sigma^2_{sys}\lesssim 10^{-7}$ would ask $\sigma_{R^2}/R^2\lesssim 10^{-3}$ and $\sigma_e\lesssim10^{-3}$. This requirement will be a big challenges for the PSF reconstruction methods using PCA. Moffatlets fit the requirement very well but just because no sharp spikes presented in the data. It's an issue to model sharp spikes using circular symmetric Moffatlets.

\begin{table}[!hbp]
\begin{tabular}{|c|c|c|c|c|}
\hline
  & 10 & 40 & 80 & ran(10-100) \\
\hline
R02S01-diffraction-OFF    & 0.0080  &  0.0148  &  0.0280  &  0.0209 \\
\hline
R02S01-diffraction-ON    & 0.0079  &  0.0136  &  0.0223  &  0.0150 \\
\hline
R22S11-diffraction-OFF    & 0.0066  &  0.0128  &  0.0234  &  0.0198 \\
\hline
R22S11-diffraction-ON    & 0.0074  &  0.0108  &  0.0201  &  0.0163 \\
\hline
\end{tabular}
\caption{Value of $[\sigma_{R^2}/R^2]$ for EMPCA in different simulation runs.}
\label{Table3}
\end{table}

\begin{table}[!hbp]
\begin{tabular}{|c|c|c|c|c|}
\hline
  & 10 & 40 & 80 & ran(10-100) \\
\hline
R02S01-diffraction-OFF    &0.0058  &  0.0090  &  0.0160  &  0.0149\\
\hline
R02S01-diffraction-ON    &0.0026  &  0.0055  &  0.0101  &  0.0081\\
\hline
R22S11-diffraction-OFF    &0.0072  &  0.0093  &  0.0147  &  0.0151\\
\hline
R22S11-diffraction-ON    &0.0024  &  0.0048  &  0.0087  &  0.0075\\
\hline
\end{tabular}
\caption{Value of $[\sigma_{R^2}/R^2]$ for Moffatlets in different simulation runs.}
\label{Table4}
\end{table}

\begin{table}[!hbp]
\begin{tabular}{|c|c|c|c|c|}
\hline
  & 10 & 40 & 80 & ran(10-100) \\
\hline
R02S01-diffraction-OFF    &0.0635  &  0.0636  &  0.0647  &  0.0672\\
\hline
R02S01-diffraction-ON    &0.0103  &  0.0111  &  0.0134  &  0.0142\\
\hline
R22S11-diffraction-OFF    &0.0578  &  0.0576  &  0.0581  &  0.0611\\
\hline
R22S11-diffraction-ON    &0.0100  &  0.0106  &  0.0123  &  0.0130\\
\hline
\end{tabular}
\caption{Value of $[\sigma_{R^2}/R^2]$ for Gaussianlets in different simulation runs.}
\label{Table5}
\end{table}

\begin{table}[!hbp]
\begin{tabular}{|c|c|c|c|c|}
\hline
  & 10 & 40 & 80 & ran(10-100) \\
\hline
R02S01-diffraction-OFF    &0.0018  &  0.0029   & 0.0037  &  0.0028\\
\hline
R02S01-diffraction-ON    &0.0018  &  0.0025  &  0.0037  &  0.0035\\
\hline
R22S11-diffraction-OFF    &0.0013  &  0.0017  &  0.0030  &  0.0031\\
\hline
R22S11-diffraction-ON    &0.0016  &  0.0027  &  0.0043  &  0.0040\\
\hline
\end{tabular}
\caption{Value of $\sigma_e$ for EMPCA method in different simulation runs.}
\label{Table6}
\end{table}

\begin{table}[!hbp]
\begin{tabular}{|c|c|c|c|c|}
\hline
  & 10 & 40 & 80 & ran(10-100) \\
\hline
R02S01-diffraction-OFF    &0.0009  &  0.0010  &  0.0013   & 0.0012\\
\hline
R02S01-diffraction-ON    &0.0004  &  0.0004  &  0.0005   & 0.0005\\
\hline
R22S11-diffraction-OFF    &0.0006  &  0.0007 &   0.0009 &   0.0009\\
\hline
R22S11-diffraction-ON    &0.0004   & 0.0004  &  0.0005  &  0.0005\\
\hline
\end{tabular}
\caption{Value of $\sigma_e$ for Moffatlets/Gaussianlets in different simulation runs.}
\label{Table7}
\end{table}

% Authors can give a citation as `\citealt{Michel+etal+1992}'.
% You may also use \cite, \citep and \citet for citation, and use Table~1
% or Figure~1 and so forth. Using \ref and \label for cross-references of
% Tables/Figures is a good way in adjusting/adding/removing text, tables or

\section{Conclusions and Discussion}
\label{sect:5}
We use three methods to reconstruct the simulated star images. The basis function methods use smooth functions which have explicit formulas and easily to be created. As our test results have shown, Moffatlets performed better in image reconstructions than Gaussianlets. This is mainly  because the Moffat function is a better PSF model than the Gaussian function.

Due to the pixelization and finite size, our basis functions are not exactly orthogonal to each other. This will rise cross-correlation between the coefficients. Numerical othogoalization algorithms, such as Gram-Schmidt process or Householder transformation can be applied to resolve this problem. The basis function methods use circular symmetric functions which are then shaped into elliptical ones according to the premeasured ellipticity of stars. As a result, they only consider the radial variation. High-order angular structures, such as diffraction spikes, usually appear in the realistic PSFs and will introduce bias into our results in principle. Our tests show that this is not a big issue in our simulated LSST images since the presented diffraction spikes are not very sharp and Moffatlets can reconstruct very well at least in terms of ellipticity and size of PSF. The more detailed studies on the issue of sharp diffraction spikes is beyond the scope of this paper.

The EMPCA method has several advantages than the basis function method: 1)The resolved PCs are compact and flexible. It's the most efficient way to reconstruct the irregular images. 2)The PCs are orthogonal to each other, this makes their coefficients independent and can be interpolated easily. While PCA method also has defects, the resolved PCs contain noise inevitably. This would introduce relatively higher scatter in the size and ellipticity in the reconstructed PSF and then rise higher systematic bias in the cosmic shear measurement.

The current PSF reconstructions is still not accurate enough to produce satisfying shear measurement for weak lensing surveys. The high-order angular structures like diffraction spikes has not been considered in detail in this paper. Meanwhile our tests have shown that the Moffatelets is a very promising tool for the PSF reconstruction. In the future studies, we will combine the angular structures of Moffatlets and EMPCA techniques together to search PCs in a finite Moffatlets space.

\section*{Acknowledgements}
BS and GL are supported by the National Key Basic Research Program of China (2015CB857000), the ¡°Strategic Priority Research Program the Emergence of Cosmological Structures¡± of the Chinese Academy of Sciences (XDB09000000). GL also thanks the supports from the One-Hundred-Talent fellowships of CAS and the NSFC grants (11273061 and 11333008). BS acknowledge the support from the NSFC grant (11403103). JC, JRP and WC acknowledge the supports from Purdue University, the Department of Energy (DE-SC00099223), and the LSST Project (C44054L).
\normalem
%\begin{acknowledgements}

%\end{acknowledgements}

\appendix
\section{Mathematical derivation of radial functions of Moffatlets}
\label{app:1}
In this appendix, we give the derivation of the radial fuctions of the mofattlets, showing that the basis functions have an analytic form. We require the radial basis functions are modified by a weight function with moffat profile
\begin{equation}
  Q_l(r)=R_l(r)w(r),w(r)=[1+(\frac{r}{r_d})^2]^{-\beta}.
  \label{eq:A1}
\end{equation}
The radial basis functions satisfy
\begin{equation}
  2\pi\int_{0}^{+\infty}dr rR_l(r)R_{l'}(r)e^{-u(r)}=\delta_{ll'},
  \label{eq:A2}
\end{equation}
where
\begin{equation}
  u(r)=-ln[[1+(\frac{r}{r_d})^2]^{-2\beta}].
\end{equation}
The corresponding  invers function is
\begin{equation}
r=a(u)=r_d\sqrt{e^{u/2\beta}-1}
\end{equation}
And we have
\begin{equation}
  rdr=\frac{r_d^2}{4\beta}e^{\frac{u}{2\beta}}du,
  \label{eq:A3}
\end{equation}

Then Eq.\ref{eq:A2} reads
\begin{equation}
 \frac{\pi r_d^2}{2\beta}\int_{0}^{+\infty}du R_{l}[a(u)]R_{l'}[a(u)]e^{-(1-1/2\beta)u}=\delta_{ll'}.
 \label{eq:A4}
\end{equation}
We change the variables according to $v(u)=(1-1/2\beta)u$ and introduce  a new  function to relate $r$ and $v$ as $r=b(v).$ Eq.\ref{eq:A2} reads
\begin{equation}
 \frac{\pi r_d^2}{2\beta-1}\int_{0}^{+\infty}dv R_{l}[b(v)]R_{l'}[b(v)]e^{-v}=\delta_{ll'}.
 \label{eq:A5}
\end{equation}
We noticed that this equation is similar with weighted integration of Laguerre ploynomials (Eq.\ref{eq:3})
\begin{equation}
 \int_{0}^{+\infty}dv L_{l}(v)L_{l'}(v)e^{-v}=\delta_{ll'}.
 \label{eq:A6}
\end{equation}
Then we have
\begin{equation}
 R_l(r)=\sqrt{\frac{2\beta-1}{\pi r_d^2}}L_l[v(r)],
 \label{eq:A7}
\end{equation}
where
\begin{equation}
   v(r)=(\frac{1}{2\beta}-1)ln[1+(\frac{r}{r_d})^2]^{-2\beta}.
\end{equation}
The final radial basis function reads
\begin{equation}
   Q_l(r)=\sqrt{\frac{2\beta-1}{\pi r^2_d}}L_l[v(r)][1+(\frac{r}{r_d})^2]^{-\beta}.
\label{eq:A8}
\end{equation}


\begin{thebibliography}{}
\expandafter\ifx\csname natexlab\endcsname\relax\def\natexlab#1{#1}\fi

\bibitem[Bartelmann \& Schneider(2001)]{2001PhR...340..291B} Bartelmann, M., \& Schneider, P.\ 2001,  340, 291

\bibitem[Alcock et al.(2000)]{2000ApJ...542..281A} Alcock, C., Allsman, R.~A., Alves, D.~R., et al.\ 2000, 542, 281

\bibitem[Cooray \& Sheth(2002)]{2002PhR...372....1C} Cooray, A., \& Sheth, R.\ 2002, 372, 1

\bibitem[Mellier(1999)]{1999ARA&A..37..127M} Mellier, Y.\ 1999, 37, 127

\bibitem[Refregier(2003)]{2003ARA&A..41..645R} Refregier, A.\ 2003, 41, 645

\bibitem[Van Waerbeke et al.(2013)]{2013MNRAS.433.3373V} Van Waerbeke, L., Benjamin, J., Erben, T., et al.\ 2013, 433, 3373

\bibitem[Planck Collaboration et al.(2015)]{2015arXiv150201589P} Planck Collaboration, Ade, P.~A.~R., Aghanim, N., et al.\ 2015, arXiv:1502.01589

\bibitem[Persic et al.(1996)]{1996MNRAS.281...27P} Persic, M., Salucci, P., \& Stel, F.\ 1996, 281, 27

\bibitem[Riess et al.(2004)]{2004ApJ...607..665R} Riess, A.~G., Strolger, L.-G., Tonry, J., et al.\ 2004, 607, 665

\bibitem[Moffat(2006)]{2006astro.ph..8675M} Moffat, J.~W.\ 2006, arXiv:astro-ph/0608675

\bibitem[Hoekstra \& Jain(2008)]{2008ARNPS..58...99H} Hoekstra, H., \& Jain, B.\ 2008, Annual Review of Nuclear and Particle Science, 58, 99

\bibitem[Massey et al.(2010)]{2010RPPh...73h6901M} Massey, R., Kitching, T., \& Richard, J.\ 2010, Reports on Progress in Physics, 73, 086901

\bibitem[Huterer(2010)]{2010GReGr..42.2177H} Huterer, D.\ 2010, General Relativity and Gravitation, 42, 2177

\bibitem[Laureijs et al.(2011)]{2011arXiv1110.3193L} Laureijs, R., Amiaux, J., Arduini, S., et al.\ 2011, arXiv:1110.3193

\bibitem[LSST Science Collaboration et al.(2009)]{2009arXiv0912.0201L} LSST Science Collaboration, Abell, P.~A., Allison, J., et al.\ 2009, arXiv:0912.0201

\bibitem[Spergel et al.(2015)]{2015arXiv150303757S} Spergel, D., Gehrels, N., Baltay, C., et al.\ 2015, arXiv:1503.03757

\bibitem[Kaiser et al.(1995)]{1995ApJ...449..460K} Kaiser, N., Squires, G., \& Broadhurst, T.\ 1995, 449, 460

\bibitem[Luppino \& Kaiser(1997)]{1997ApJ...475...20L} Luppino, G.~A., \& Kaiser, N.\ 1997, 475, 20

\bibitem[Hoekstra et al.(1998)]{1998ApJ...504..636H} Hoekstra, H., Franx, M., Kuijken, K., \& Squires, G.\ 1998, 504, 636

\bibitem[Refregier \& Bacon(2003)]{2003MNRAS.338...48R} Refregier, A., \& Bacon, D.\ 2003, 338, 48

\bibitem[Bridle et al.(2010)]{2010MNRAS.405.2044B} Bridle, S., Balan, S.~T., Bethge, M., et al.\ 2010,405, 2044

\bibitem[Kitching et al.(2012)]{2012MNRAS.423.3163K} Kitching, T.~D., Balan, S.~T., Bridle, S., et al.\ 2012, 423, 3163

\bibitem[Kitching et al.(2013)]{2013ApJS..205...12K} Kitching, T.~D., Rowe, B., Gill, M., et al.\ 2013,  205, 12

\bibitem[Mandelbaum et al.(2014)]{2014ApJS..212....5M} Mandelbaum, R., Rowe, B., Bosch, J., et al.\ 2014, 212, 5

\bibitem[Paulin-Henriksson et al.(2008)]{2008A&A...484...67P} Paulin-Henriksson, S., Amara, A., Voigt, L., Refregier, A., \& Bridle, S.~L.\ 2008, 484, 67

\bibitem[Paulin-Henriksson et al.(2009)]{2009A&A...500..647P} Paulin-Henriksson, S., Refregier, A., \& Amara, A.\ 2009, 500, 647

\bibitem[Massey et al.(2013)]{2013MNRAS.429..661M} Massey, R., Hoekstra, H., Kitching, T., et al.\ 2013, 429, 661

\bibitem[Berg{\'e} et al.(2012)]{2012MNRAS.419.2356B} Berg{\'e}, J., Price, S., Amara, A., \& Rhodes, J.\ 2012, 419, 2356

\bibitem[Massey \& Refregier(2005)]{2005MNRAS.363..197M} Massey, R., \& Refregier, A.\ 2005, 363, 197

\bibitem[Refregier(2003)]{2003MNRAS.338...35R} Refregier, A.\ 2003, 338, 35

\bibitem[Li et al.(2013)]{2013IAUS..288..306L} Li, G., Xin, B., \& Cui, W.\ 2013, Astrophysics from Antarctica, 288, 306

\bibitem[Bailey(2012)]{2012PASP..124.1015B} Bailey, S.\ 2012, 124, 1015

\bibitem[Saglia et al.(1993)]{1993MNRAS.264..961S} Saglia, R.~P., Bertschinger, E., Baggley, G., et al.\ 1993, 264, 961

\bibitem[Andrae et al.(2011)]{2011MNRAS.417.2465A} Andrae, R., Melchior, P., \& Jahnke, K.\ 2011, 417, 2465

\bibitem[Li et al.(2012)]{2012arXiv1203.0571L} Li, G., Xin, B., \& Cui, W.\ 2012, arXiv:1203.0571

\bibitem[Shlens(2014)]{2014arXiv1404.1100S} Shlens, J.\ 2014, arXiv:1404.1100

\bibitem[Kitching et al.(2010)]{2010arXiv1009.0779K} Kitching, T., Balan, S., Bernstein, G., et al.\ 2010, arXiv:1009.0779

\bibitem[Trujillo et al.(2001)]{2001MNRAS.328..977T} Trujillo, I., Aguerri, J.~A.~L., Cepa, J., \& Guti{\'e}rrez, C.~M.\ 2001, 328, 977

\bibitem[Hu(1999)]{1999ApJ...522L..21H} Hu, W.\ 1999, 522, L21

\bibitem[Peterson et al.(2015)]{2015ApJS..218...14P} Peterson, J.~R., Jernigan, J.~G., Kahn, S.~M., et al.\ 2015, 218, 14


\end{thebibliography}
\end{document}